\documentclass[twocolumn]{aastex63}

\usepackage{amsmath}

\shorttitle{V892 Tau}
\shortauthors{Vides}

\graphicspath{{./}{figures/}}
\begin{document}

\title{High Angular Resolution Imaging of the V892 Tau Binary System: A New Circumprimary Disk Detection and Updated Orbital Constraints}

\author{Christina L. Vides}
\author{Steph Sallum}
\affiliation{Department of Physics and Astronomy, University Of California, Irvine}
\author{Josh Eisner}
\affiliation{Department of Astronomy and Steward Observatory, University of Arizona}
\author{Andy Skemer}
\author{Ruth Murray‑Clay}
\affiliation{ Astronomy \& Astrophysics Department, University of California, Santa Cruz}

\begin{abstract}
\par We present a direct imaging study of V892 Tau, a young Herbig Ae/Be star with a close-in stellar companion and circumbinary disk. Our observations consist of images acquired via Keck 2/NIRC2 with non-redundant masking and the pyramid wavefront sensor at K$^\prime$ band (2.12$\mu$m)  and L$^\prime$ band (3.78$\mu$m).
Sensitivity to low-mass accreting companions and cool disk material is high at L$^\prime$ band, while complimentary observations at K$^\prime$ band probe hotter material with higher angular resolution. These multi-wavelength, multi-epoch data allow us to differentiate the secondary stellar emission from disk emission and deeply probe the structure of the circumbinary disk at small angular separations. We constrain architectural properties of the system by fitting geometric disk and companion models to the K$^\prime$ and L$^\prime$ band data. From these models, we constrain the astrometric and photometric properties of the stellar binary and update the orbit, placing the tightest estimates to date on the V892 Tau orbital parameters. We also constrain the geometric structure of the circumbinary disk, and resolve a circumprimary disk for the first time.

\end{abstract}

\keywords{Near-Infrared Direct Imaging --- 
Transition Disks --- Circumbinary Disks}

\section{Introduction} \label{sec:intro}
\par 
Circumbinary planets have only recently been observed \citep[e.g.][]{2003Sci...301..193S,2011Sci...333.1602D}, despite long-standing predictions of planets existing in stable orbits around stellar binaries \citep[e.g.][]{1980CeMec..22....7S,1988A&A...191..385R,1993CeMDA..56...45B}. Out of the $\sim$5000 confirmed exoplanets, only ~$\sim$75 have been confirmed to be in a circumbinary (CB) orbit, an orbital path where a planet orbits both the primary and secondary stars \citep[e.g.][]{2013PASP..125..989A}. These mature CB planets are an enigma for planet formation theory, since their semi-major axes are close-in, near the limits of instability \citep{2011Sci...333.1602D,2020AJ....159...94S}. These close-in separations suggest that migration is one of the key mechanisms in CB planet formation, but many aspects of this process remain untested \citep[e.g][]{2006ApJ...642..478M,2008A&A...483..633P,2014A&A...564A..72K,2021A&A...645A..68P,2022MNRAS.513.2563C}. Understanding the details of the locations and the timescales on which CB planets form requires better constraints on the youngest of these systems where planets are actively forming.  

High angular resolution studies of circumbinary disks enable us to characterize the initial conditions of CB planet formation. Constraining their dust distributions can advance our understanding of where planet formation may be ongoing \citep[e.g. in reservoirs of circumbinary, circumprimary, or circumsecondary material][]{2012A&A...539A..18M,2015A&A...582A...5L}. Such studies can also inform our understanding of migration mechanisms and how they are influenced by the spatial properties of the CB disk \citep{2008A&A...478L..31G,2009ApJ...700..491M,2017ApJ...840...60B,2018ApJ...869L..44K}. Furthermore, theoretical predictions of protoplanet spectral energy distributions (SEDs) suggest that they have low contrasts at infrared (IR) wavelengths \citep[e.g.][]{2015ApJ...799...16Z,2015ApJ...803L...4E,2003A&A...402..701B}. Hence, in addition to disk characterization, searching for and characterizing actively-forming or recently-formed planets in the IR would directly constrain the initial conditions of CB planet formation.
 
\subsection{V892 Tau} \label{sec:v892tau}
One natural laboratory for studying circumbinary planet formation is V892 Tau, a young Herbig Ae/Be star located at a distance of $\sim$135 pc within the Taurus-Auriga star-forming region \citep[][]{2020yCat.1350....0G}.  Spectral type estimates of the primary vary from A0 to B9 \citep[][]{1992ApJ...397..613H,1994ApJ...424..237S,2004AJ....127.1682H,2014ApJ...786...97H}. When imaged at near-IR wavelengths, V892 Tau is shown to host an almost equally bright secondary companion at angular separations ranging between 40-65 mas, with the most recent constraints on the orbital parameters being: semi-major axis a = 7.1 $\pm$ 0.1 au, period P = 7.7 $\pm$ 0.2 yr, eccentricity \(e\) = 0.27 $\pm$ 0.1 and inclination $i$ = 59.3 $\pm$ \(2.7^{\circ}\) \citep[][]{2005A&A...431..307S,2008ApJ...681L..97M,2021ApJ...915..131L}. From CO Keplerian gas rotation, the total mass of the system is determined to be 6.0 $\pm$ 0.2 \(M_\odot\) \citep[][]{2021ApJ...915..131L}.

\par The circumstellar environment of V892 Tau and its companion has also been well studied in both the mid-infrared and the millimeter continuum. The circumbinary disk was first discovered when imaged at 10.7 $\micron$. An elongated structure with two bright lobes was detected and fit with a 2-dimensional Gaussian model to estimate the inclination of the disk \citep[][]{2008ApJ...681L..97M}. In the millimeter continuum,  V892 Tau has a radially asymmetric dust ring with peak mm emission at 0.2$^{\prime\prime}$ and enough mass to form giant planets \citep[][]{2021ApJ...915..131L}. 
Warping in the CB disk has been tentatively identified in mm and 10.7 $\micron$ imaging \citep[][]{2008ApJ...681L..97M,2021ApJ...915..131L}. 
This scenario is consistent with the V892 Tau binary eccentricity and mass ratio, since highly eccentric orbits are known to induce warping and tidal truncation within the circumbinary disks of near equal mass binaries \citep[][]{1994ApJ...421..651A,2020MNRAS.498.2936H,2015MNRAS.452.2396M}.
%The V892 Tau stellar companion mass is nearly equal to the mass of the primary, but highly eccentric orbits are known to induced warping and tidal truncation within the circumbinary disks of near equal mass binaries \citep[][]{1994ApJ...421..651A,2020MNRAS.498.2936H,2015MNRAS.452.2396M}. Such warping has been tentatively identified in mm and 10.7 $\mircon$ imaging \citep[][]{2008ApJ...681L..97M,2021ApJ...915..131L}. 
In addition to characterizing the CB disk, mid-IR long baseline interferometry has tentatively suggested the presence of a resolved circumprimary disk, but was unable to distinguish a circumprimary disk from an additional dusty companion \citep[][]{2019PhDT.......134C}.

\subsection{Paper Outline}
We present 2-4 micron high angular resolution observations of the V892 Tau system. The paper is formatted as follows: Section \ref{sec:methods} describes the observations and data reduction. Section \ref{sec:analysis} describes the image reconstruction method and analytical model fitting to the data. In Section \ref{sec:results}, we statistically compare the results of each model and the correlations between the models and the data; then we place constraints on the geometry of the system. We then append our results to prior astrometric data and fit an orbit to the V892 Tau stellar companion. In Section \ref{sec:discussion}, we discuss the implications of the results and estimate our sensitivity to additional, planetary-mass companions in the system. We conclude in Section \ref{sec:conclusion}.

 \section{Methods}\label{sec:methods}

 \subsection{Non-Redundant Masking}\label{NRM}
Nearby star-forming regions are located at distances $\gtrsim$ 100 pc, where the angular resolution provided by traditional direct imaging methods is insufficient for IR planet searches on orbital scales of $\lesssim~10-15$ AU \citep[e.g.][]{2013ApJ...767...11G}. Resolving smaller scales at such distances requires interferometric techniques. One method is non-redundant masking (NRM), an aperture masking technique in which a conventional telescope is turned into an interferometric array by placing a mask with discrete holes in the pupil plane \citep[e.g.][]{2000PASP..112..555T}. Each pair of holes, also known as a baseline, has a distinct orientation and separation, such that each baseline has its own unique spatial frequency (hence the term non-redundant). 
 
\par The image on the detector, or the interferogram, shows the interference fringes formed by the mask. We take the Fourier transform of the interferogram to get the complex visibilites (which have the form \(Ae^{i\phi}\)). The complex visibility for each baseline is located in an extended region in Fourier space due to the finite size of the holes and the wavelength coverage set by the observing bandpass. From the appropriate regions in Fourier space, we calculate two quantities: squared visibilities and closure phases. Squared visibilities are the squares of the complex visibility amplitudes, and give the power corresponding to each baseline \citep[e.g.][]{1958MNRAS.118..276J}. Closure phases are sums of phases around baselines that form a triangle \citep[e.g.][]{1986Natur.320..595B}. Closure phases are highly sensitive to asymmetries and cancel first order wavefront errors, leaving only intrinsic phase and higher order residual errors. Closure phases and squared visibilities can be used to understand the source brightness distribution via both model fitting and image reconstruction. 

\par NRM allows for moderate contrast ($\sim$1:100-1:1000) at smaller angular separations ($\gtrsim$ 0.5$\lambda$/D) than those probed by traditional imaging techniques such as coronagraphy \citep[e.g.][]{2012SPIE.8442E..04M,2006ApJS..167...81G,2019SPIE11117E..1FR,2019JATIS...5a8001S}. Observations using NRM have been successful in probing close-in protoplanetary disk structures \citep[e.g.][]{2019ApJ...883..100S} and identifying companions \citep[e.g.][]{2008ApJ...678L..59I}. Here we apply NRM on Keck 2/NIRC2 to deeply probe the structure of the V892 Tau circumbinary disk.

\subsection{Observations}\label{obs}
 We used the 9-hole NRM in Keck 2/NIRC2 in conjunction with the Pyramid Wavefront Sensor (PyWFS) to directly image V892 Tau with the L$'$ filter (central \(\lambda\) = 3.776 $\micron$) and K$'$ filter (central \(\lambda\) = 2.124 $\micron$). Observations took place from 10:06 UT until 15:37 UT on November 6th, 2020 (L$'$) and at 4:57 UT until 10:33 UT on January 21st, 2022 (K$'$). The median seeing for the first half-night of November 5th, 2020 was 0.76$^{\prime\prime}$, with a minimum of 0.49$^{\prime\prime}$, a maximum of 1.26$^{\prime\prime}$, and a standard deviation of 0.16$^{\prime\prime}$ as measured by the Differential Image Motion Monitor. The median seeing for the first half-night of January 20th, 2022 was 1.11$^{\prime\prime}$, with a minimum of 0.75$^{\prime\prime}$, a maximum of 1.95$^{\prime\prime}$, and a standard deviation of 0.3$^{\prime\prime}$.

 \par
\begin{figure*}[ht!]
\plottwo{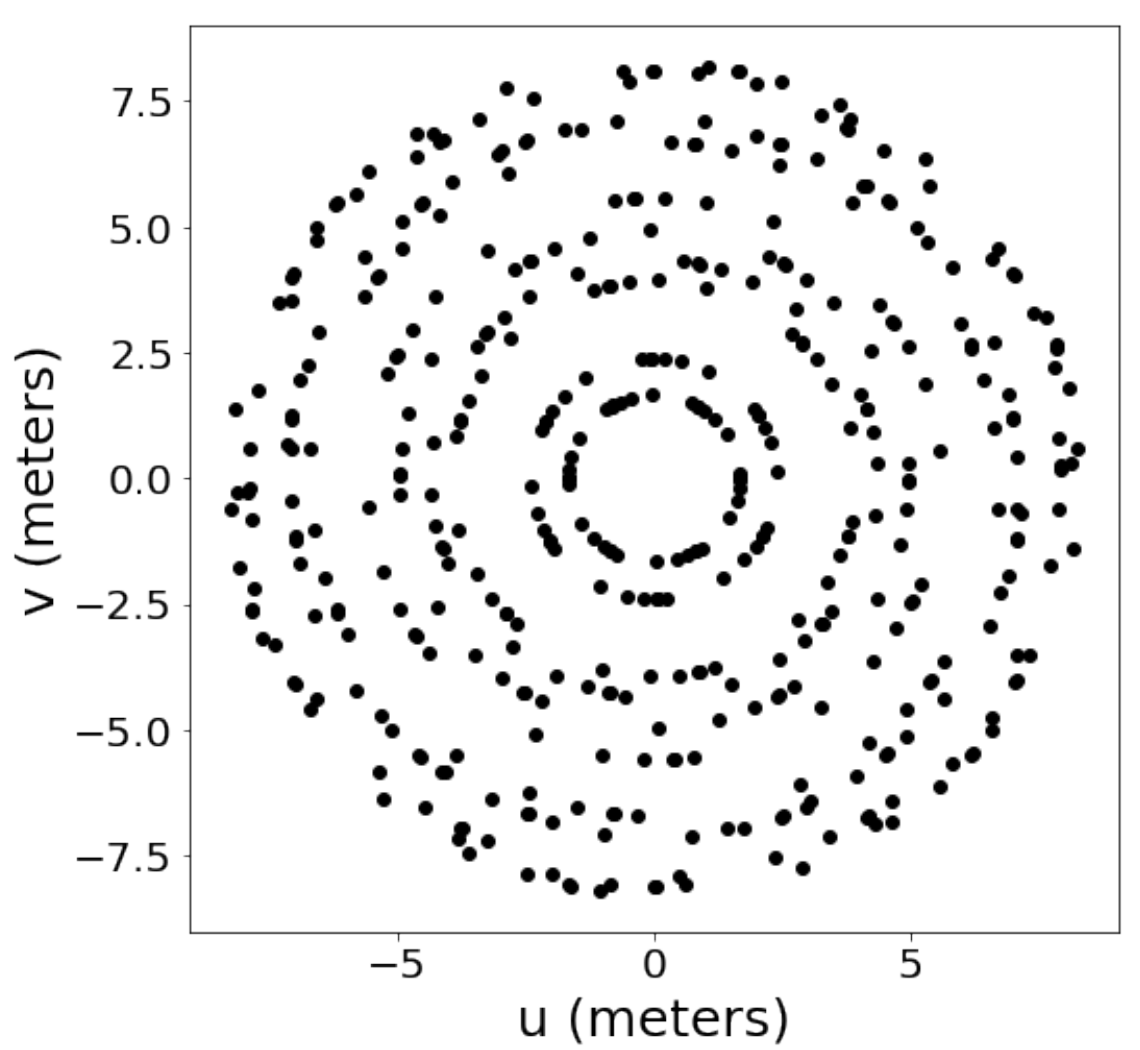}{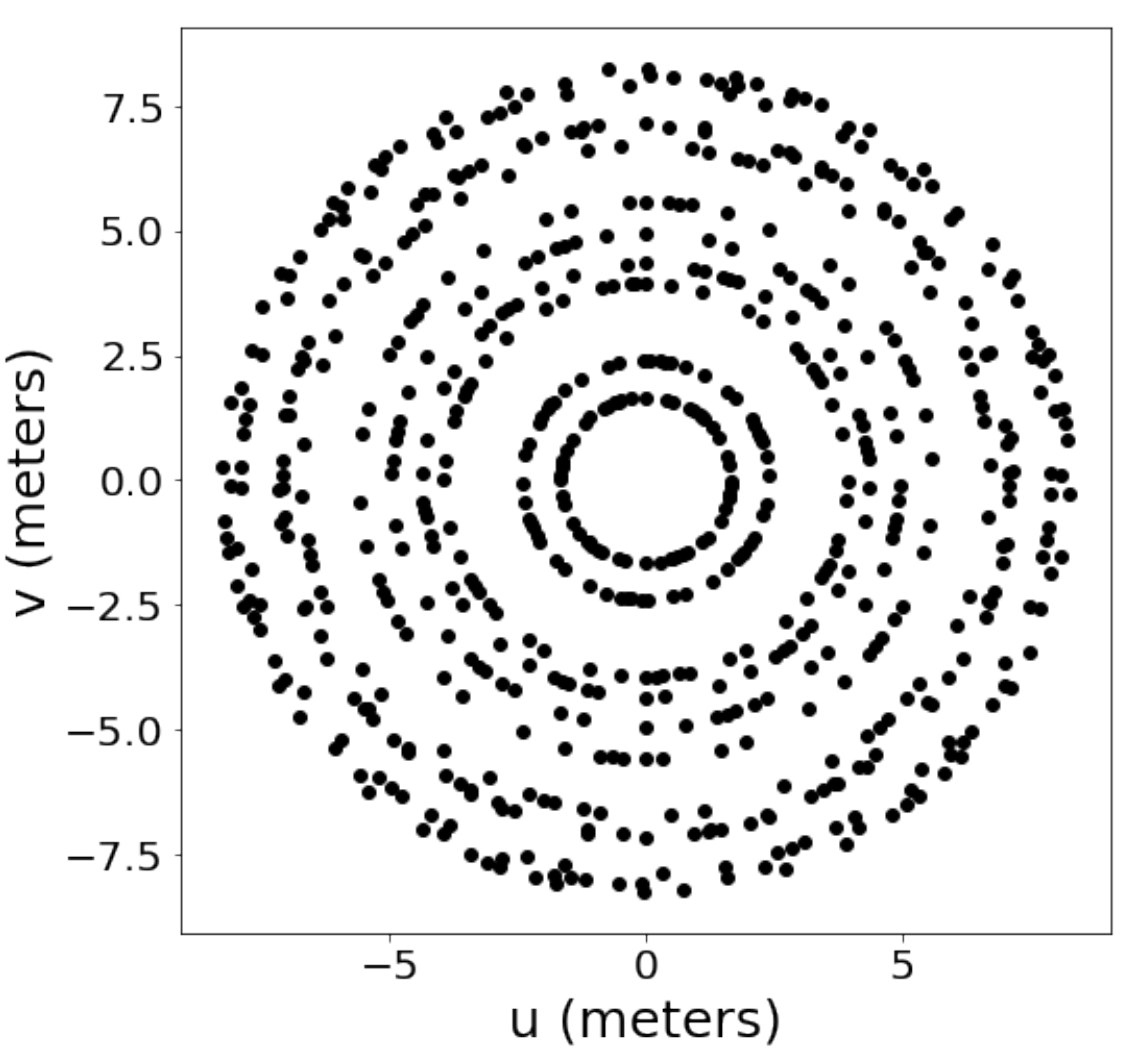}
\caption{Scattered points show uv-coverage (the rotation of each baseline with parallactic angle) at L$^\prime$ band (left) and K$^\prime$ band (right). We obtained $\sim$64$^\circ$ and $\sim$108$^\circ$ of sky rotation at L$^\prime$ and K$^\prime$, respectively.}\label{fig:uv}
\end{figure*}

 At both wavelengths, we observed V892 Tau for a half-night centered on transit. We observed in vertical angle mode, which allows baselines to rotate on the sky. This fills the Fourier plane and allows for astrophysical signals to rotate while instrumental systematics remain fixed, enabling angular differential calibration. Figure \ref{fig:uv} shows the rotation of each baseline with parallactic angle for the duration of each half-night at both wavelengths. We obtained $\sim$64 degrees and $\sim$108 degrees of sky rotation at L$^\prime$ and K$^\prime$, respectively.
 
\par During each night we alternated between observing the science target and point spread function (PSF) calibrators, which are used to estimate higher order wavefront errors. To choose appropriate calibrators, we optimized between matching WFS brightnesses for similar quality adaptive optics (AO) correction, brightnesses at the science wavelength for efficient integration times, and separations on the sky to maximize common atmospheric paths and minimize slew times. 
%However, it is important that the wavefront sensing flux is well-matched for a similar quality AO correction and to achieve roughly the same high order instrumental errors. The frame rate of the WFS should be constant across the the science target and calibrators to ensure measurements of the same systematics. 
We chose HD 283520, HD 281928 and HD 283577, whose coordinates and fluxes relative to V892 Tau are listed in Table \ref{table:1}. The calibrators have similar brightnesses to V892 Tau in the science bandpasses, allowing for efficient integration times. 
Despite the calibrators' brighter fluxes at H band (the PyWFS bandpass), the WFS frame rate for all objects was 1054 Hz, resulting in similar quality AO correction.
The close angular separations between V892 Tau and the calibrators minimize slew overheads as well as calibration errors caused by differential refraction \citep[][]{2013MNRAS.433.1718I}.

\par We subframed the 1024 $\times$ 1024 pixel detector to 512 pixels on each side and dithered on the detector, taking 10 frames in the top-left and bottom-right corners for each pointing to enable background subtraction. We spent equal amounts of time on the science target and PSF calibrators and alternated between dither-pair sequences. The coadds and integration times were chosen to build signal to noise with the detector in a linear response regime with minimal readout overheads. We obtained $\sim$40 min and $\sim$30 min of total integration time at L$^\prime$ and K$^\prime$ band, respectively.

%% An example table using AASTeX's deluxetable. Note that since
%% only one figure OR one table is allowed this is commented out.
\begin{deluxetable*}{cccccc}
\tablecaption{Properties of V892 Tau and PSF calibrators \label{table:1}
} 
\tablehead{
\colhead{Target}  & \colhead{RA} & \colhead{Dec} & \colhead{L$'$ (Jy)} & \colhead{K$'$ (Jy)}&\colhead{H (Jy)}}
\startdata
V892 Tau & 04 18 40.61&  +28 19 15.62 & 1.75&3.23 &1.68\\
HD 283520 &04 18 27.10 &+83 41 49.30 &1.87 &4.41 &3.54\\
HD 281928& 04 20 11.51& +29 13 22.99&1.66 &3.64  &4.78\\
HD 283577 &04 21 49.04 &+27 17 04.94&1.05 &2.22&3.02
\enddata
\end{deluxetable*}  

\begin{deluxetable*}{cccccccc}
\tablecaption{Observational setup for V892 Tau and PSF calibrators at both wavelengths}
\tablehead{
\colhead{Target}  & \colhead{tint (L$'$)} & \colhead{tint (K$'$)} &\colhead{coadds (L$'$)}& \colhead{coadds (K$'$)} &\colhead{frames per dither}&\colhead{dithers (L$'$)}& \colhead{dithers (K$'$)}}
\startdata
V892 Tau& 0.5 & 0.5 &40& 20& 20&6& 9\\
HD 283520 & 1 & 0.5& 20&20&20 &2&3\\
HD 281928& 1 &0.5 &20&20&20&2&3\\
HD 283577& 1 &0.5 &20&20&20&2&3
\enddata
\end{deluxetable*}  

%notes about observing section

\subsection{Data Reduction}\label{sec:datared}
We use a well-tested pipeline, \texttt{SAMpy} \citep[][]{2017ApJS..233....9S,2022SPIE12183E..2MS}, to reduce the data. We first calibrate the images by flattening and correcting for bad pixels by replacing them with the mean of the adjacent pixels. The median of one dither is then subtracted from each image in the other dither position for each dither-pair sequence for sky subtraction. The images are then cropped and Fourier transformed to obtain complex visibilities. 

We crop the L$^\prime$ band images to 161 x 161 pixels and the K$^\prime$ band images to 91 x 91 pixels, then pad the images with zeros such that their sizes are 1024 x 1024 pixels before taking their Fourier transforms (FTs). We then sample the FT using all pixels that correspond to each baseline, and square the amplitude of the FT to obtain squared visibilities. The closure phases are then calculated such that the (u,v) coordinates of each closing triangle satisfy:
\begin{equation}
    (u_{1}, v_{1}) + (u_{2}, v_{2}) + (u_{3}, v_{3}) = 0.
\end{equation}
Sampling the FT at (u,v) coordinates satisfying the above equation, we calculate bispectra, which are products of the complex visibilities. We average the bispectra over multiple pixels for each baseline triangle for each frame, and then across the frames for each pointing. The phases of the bispectra are then taken to obtain the closure phases (Section \ref{NRM}). There are 36 squared visibilities and 84 closure phases calculated per pointing for the NIRC2 9-hole mask.

\par The squared visbilites and closure phases are next calibrated by fitting polynomial functions in time to the PSF calibrators. Polynomial orders are representative of instrumental noise variation, with a zeroth order polynomial indicating constant noise throughout the night and a high-order polynomial indicating high variability. We sample the polynomial function at the time of the science observations to estimate the instrumental systematics present in both the science closure phases and squared visibilities. 
To calibrate, we subtract the instrumental closure phases from the science closure phases, and divide the science squared visibilities by the instrumental squared visibilities.

We perform multiple calibrations with a variety of polynomial orders (ranging between zero and N-1, where N is the number of pointings).
For the final calibrated closure phases, we adopt the order that minimizes their scatter, corresponding to a first-order polynomial at L$^\prime$ band and a second-order polynomial at K$^\prime$ band.
For the squared visibilities, we assess the quality of the calibration by finding orders that minimize not only the scatter, but also the number of outliers with values $>1$.
A first-order polynomial satisfies these criteria at L$^\prime$ band, and a fourth-order polynomial at K$^\prime$ band.
The higher order polynomials at K$^\prime$ band suggest higher variability in the seeing, which is consistent with the seeing values reported in Section \ref{obs}.

We estimate error bars for the calibrated data by measuring the scatter of the calibrator squared visibilities and closure phases associated with each baseline and closing triangle, respectively. Rather than estimating the statistical error by measuring the standard deviation around the mean for each pointing, we estimate the calibration error  by measuring the standard deviation across all pointings. This captures variability caused by changing systematics such as quasi-static speckles. We assign the estimated calibration error as the error bar for each science target squared visibility and closure phase. 

This method is more appropriate than assigning statistical error bars, since calibration errors are the dominant error source in NRM observations \citep{2013MNRAS.433.1718I}. However, since some of the variations in the systematics are by definition removed during calibration, this method is conservative and tends to overestimate error bars. 
Lastly, we note that this approach means that error bars between baselines and closing triangles vary in a single pointing, but the error bars for all observables associated with a given baseline or closing triangle are constant across all N pointings. 
%Since the empirical calibrator scatter captures trends in the systematics that are removed during the calibration, this method ensures that error bar estimates are conservative.

\section{Anaylsis}\label{sec:analysis}
\subsection{Image Reconstruction}\label{imagerecon}
\par After the observables are calibrated, we reconstruct images of the science target with SQUEEZE \citep[e.g.][]{2010SPIE.7734E..2IB}, an algorithm that uses Markov-Chain Monte Carlo (MCMC) methods to fit a model image to the closure phases and squared visibilities.  
SQUEEZE allows for simultaneous image reconstruction and model fitting, with several analytic model components that can be included in the reconstructed images. We reconstruct two images for each dataset using two different SQUEEZE models - a single point source model, and a binary model since V892 Tau has a known stellar companion. The single point source model includes a central, unresolved delta function to represent the central star, and we allow its fractional flux to vary. The binary model includes two unresolved delta functions to represent the central star and the companion, and we allow their fractional fluxes and the companion position to vary. We run SQUEEZE in parallel tempering mode to efficiently explore the image parameter space. The images have a platescale of 5 milliarcseconds (mas) per pixel and a size of 100 pixels on each side.

%SQUEEZE fits a model image with the components of a central, unresolved star and a uniform disk. The fractional flux of the central star is the only varied parameter, while the size of the uniform disk at the reference wavelength, and the flux power law index are fixed parameters. SQUEEZE assumes an ambiguous source morphology in the images, unless a model component is added. Due to the sparsity of the Ft sampling, not all of the phase info can be recovered from the closure phases. Image reconstruction is an ill-posed problem which requires regularization, therefore additional geometric modeling is necessary to inform of the morphology of the image. \citep[][]{2017ApJS..233....9S}. 

\subsection{Geometric Models}\label{sec:geomods}

Due to the sparsity of NRM Fourier coverage and incomplete recovery of phase information, image reconstruction is an under-constrained problem. We thus fit models to the observables to understand the morphology of the system and to test the robustness of the reconstructed images \citep[][see also Section \ref{subsec:squeezemod}]{2017ApJS..233....9S}.
We explore geometric models that include disk and unresolved companion components, fitting them to the Fourier observables. 
The three classes of models that we include are: (1) companion-only, (2) disk-only, and (3) disk-plus-companion.
We fit them to each wavelength independently, since as geometric models they do not apply physically-motivated constraints on the relative fluxes of each component at K$^\prime$ and L$^\prime$ bands.

%From the best-fit models, we constrain astrometric and photometric properties of the V892 Tau disk and binary companion.  
%We fit three models the the data - companion-only, disk-only, and disk plus companion. 
\par In the companion-only model, we analytically take the FT of two delta functions representing the primary and secondary stars, which have separation (S), position angle (PA), and the height of the secondary representing the contrast (CC). We convert the contrast from magnitudes to a flux ratio and give the delta function representing the primary a height of 1 and the delta function representing the secondary a height equal to the flux ratio.  We let the separation, position angle, and contrast of the secondary vary between 0 and 500 mas, 0 and 360 degrees, and 0 and 8 magnitudes, respectively. We sample the model FTs at the locations corresponding to the mask baselines to calculate model closure phases and squared visibilities. To fully explore parameter space, we use \texttt{emcee} \citep[][]{Foreman_Mackey_2013}, an MCMC fitting package, in parallel tempering mode with 20 temperatures, 100 walkers, and 10,000 steps.

To model extended emission from the disk, we follow a procedure similar to that described in Appendix B of \cite{2021AJ....161...28S}. The brightness distribution of the disk is defined as:
\begin{equation}
I(x',y')=(1+A_{s}\cos(\phi_{s}+\phi))(I_{d}(x',y')-I_h(x',y'))
\end{equation}
where 
\begin{equation}
\begin{split}
I_{d}(x',y') = \exp\left(-\frac{x'^{2}}{2(\sigma_{x'})^{2}}-\frac{y'^{2}}{2(\sigma_{y'})^{2}}\right) \\
I_h(x',y') = \exp\left(-\frac{x'^{2}}{2(\sigma_{x'}f_{h})^{2}}-\frac{y'^{2}}{2(\sigma_{y'}f_{h})^{2}}\right) \\
\end{split}
\end{equation}
and
\begin{equation}
\begin{split}
x'=x\cos\theta-y\sin\theta \\
y'=x\sin\theta+y\sin\theta \\
\phi=\arctan(x,y) \\
\end{split}
\end{equation}

and $x$ and $y$ are locations in image space (increasing up and to the right).
Here, $\theta$ is the position angle of the disk major axis, which is measured east of north and allowed to vary from 0$^{\circ}$ to 180$^{\circ}$. 
The skew amplitude of the disk is given by $A_{s}$ and ranges from 0 to 1, and the peak skew position is given by $\phi_{s}$, which is allowed to vary between 0$^{\circ}$ and 360$^{\circ}$ measured east of north.
The minor-to-major axis ratio ($a_{ratio}$) and the full-width-half-maximum (FWHM) of the Gaussian disk along the major axis are given by:
\begin{equation}
a_{ratio} = \sigma_{y'}/\sigma_{x'}
\end{equation}
and
\begin{equation}
\mathrm{FWHM}=2\ln\sqrt{2}\sigma_{x'}.
\end{equation}
We let the FWHM vary from 0 to 500 mas and allow for a hole that occupies a fraction f$_h$ of the FWHM. A delta function with a fractional flux $b$ represents the central star. Both f$_h$ and $b$ are free parameters between 0 and 1. 
We use \texttt{emcee} to explore parameter space using the same parallel tempering settings as the companion-only model.

\par The disk-plus-companion model is a combination of the Gaussian disk and the two delta functions. We vary the fractional fluxes occupied by the disk and companion. In the general disk-plus-companion model, the disk is allowed to be interior or exterior to the companion. However, we also explore a model that forces the disk to be exterior to the companion which is discussed in more detail in Section \ref{sec:bestfit}. We use the same \texttt{emcee} parallel tempering settings as the previous two models. 

\subsection{Contrast Curve Generation}
We generate contrast curves from companion-only models to place mass constraints on additional, undetected companions. We generate a grid of evenly spaced companion models ranging 0 to 500 mas in separation, 0$^{\circ}$ to 360$^{\circ}$ in position angle, and 0 to 8 magnitudes in contrast. We then fit the companion models to the residuals between the data and the best-fit disk-plus-companion model, calculating a $\chi^{2}$ for each model. To obtain a contrast curve, we average over the grid in position angle and then calculate $\chi^{2}$ intervals between the null model (no companion) and the companion-only models at each separation and contrast.
At a given separation, we adopt the contrast with a $\chi^{2}$ interval of 25 as the 5$\sigma$ contrast \citep[e.g.][]{2019ApJ...883..100S}.
This method gives similar results as fitting the grid of companion models to a PSF calibrator that has undergone the calibration process (see Section \ref{sec:datared}).

\begin{figure*}[ht!]
\plotone{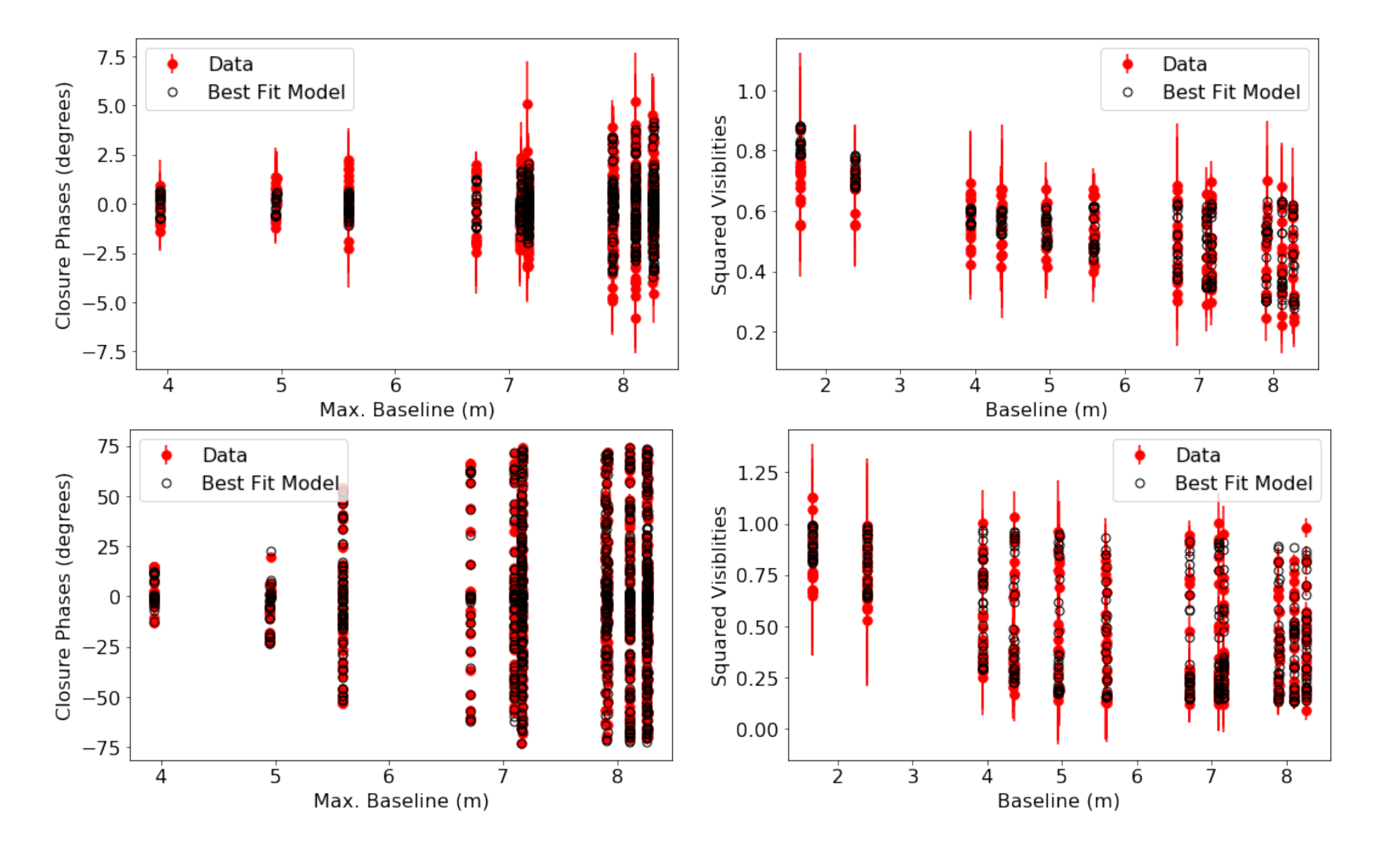}
\caption{V892 Tau Fourier observables. Left panels show closure phases for L$^\prime$ band (top) and K$^\prime$ band (bottom). The red points with error bars show the closure phases calculated from the data as described in Section \ref{sec:datared}. Hollow black circles show the closure phases calculated from the best-fit geometric model as described in Section \ref{sec:geomods}. Right panels show the squared visibilities at L$^\prime$ band (top) and K$^\prime$ band (bottom). The red points with error bars show the squared visibilites calculated from the data as described in Section \ref{sec:datared}. Hollow black circles show the squared visibilities calculated from the the best fit geometric model as described in Section \ref{sec:geomods}. }\label{fig:results}
\end{figure*}

\begin{figure*}[ht!]
\plottwo{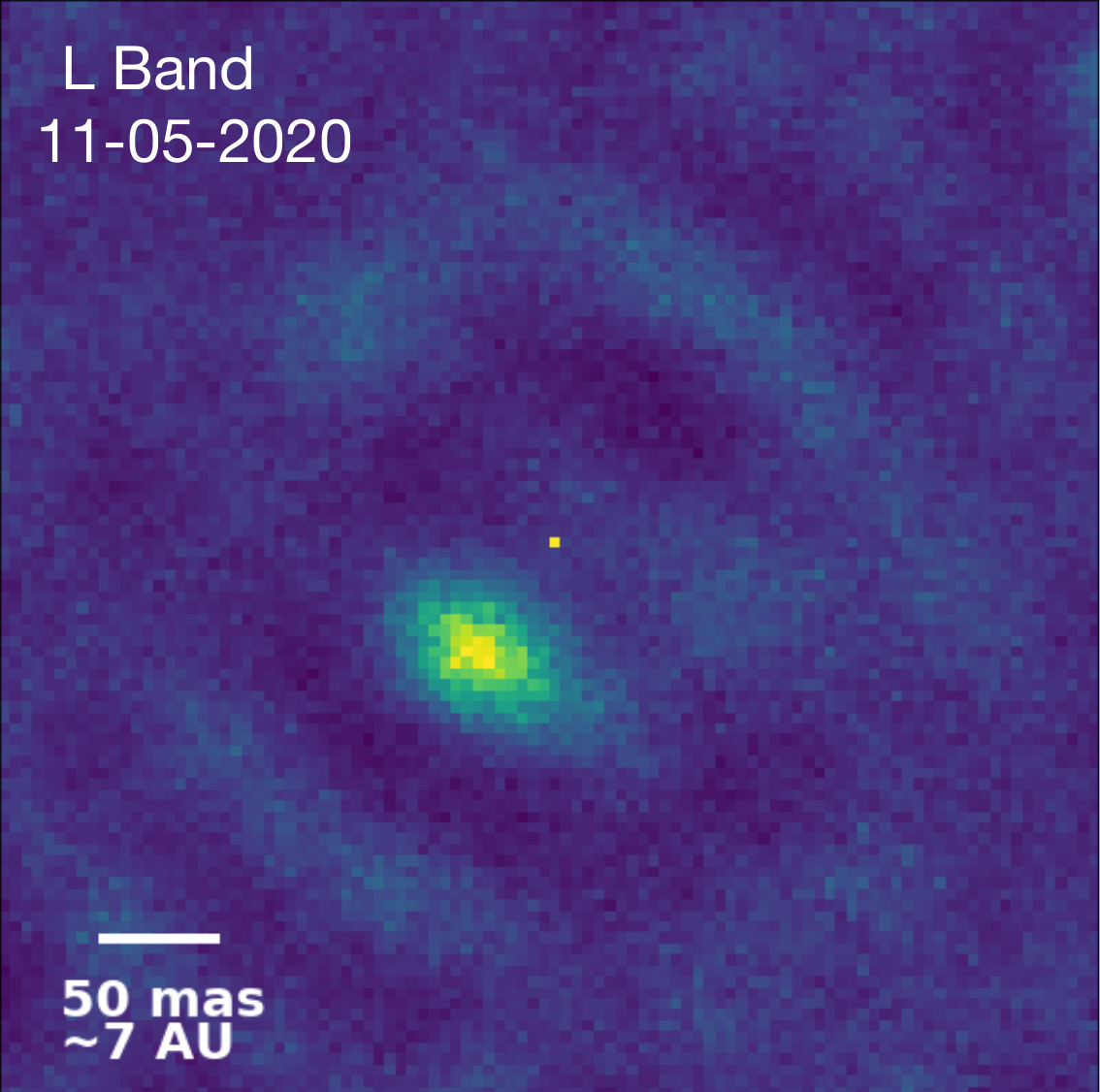}{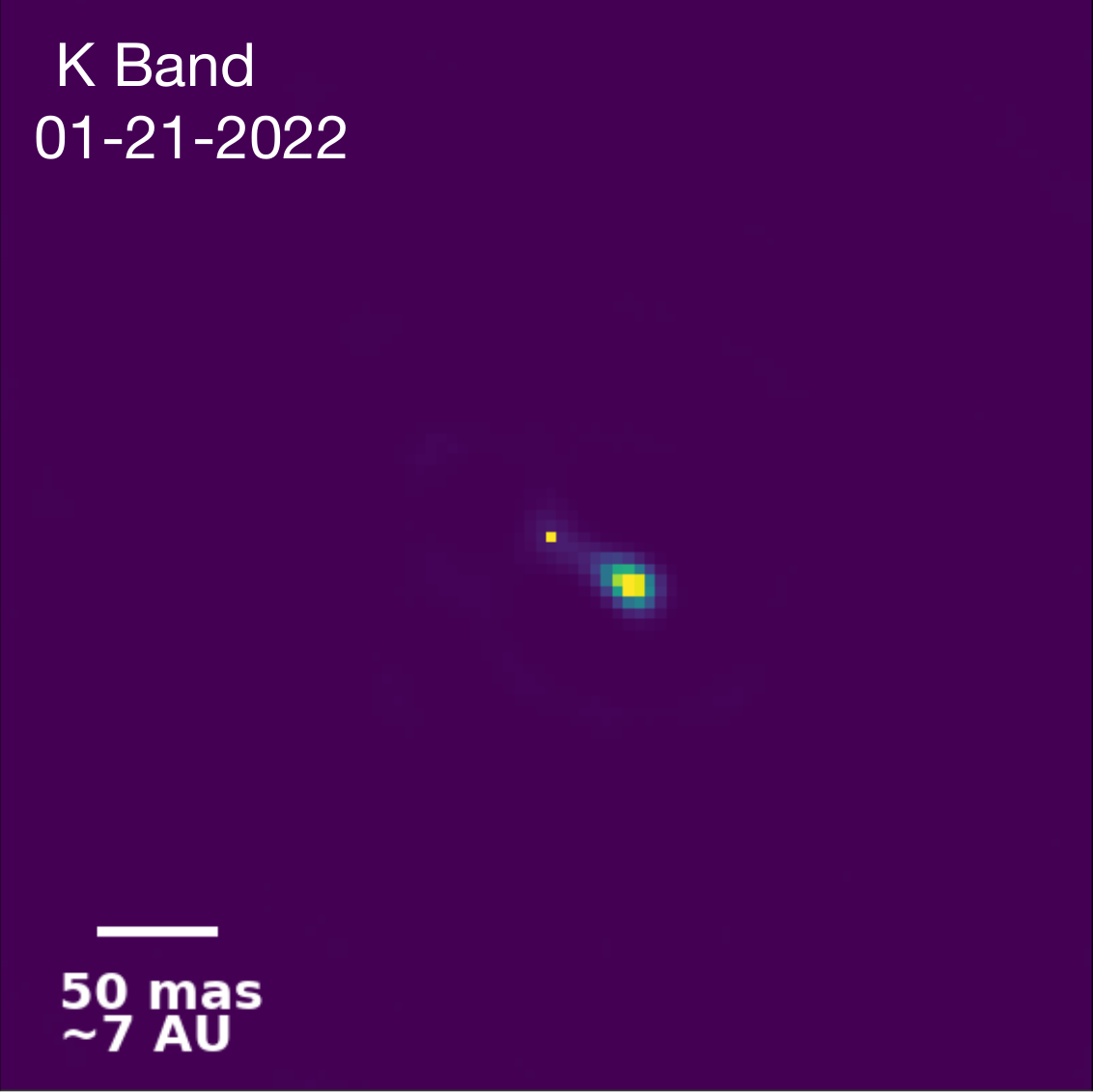}
\caption{SQUEEZE images reconstructed from the L$^\prime$ band (left) and K$^\prime$ band (right) observables. The central stars are represented with $\delta$ functions occupying fractional fluxes of 0.71 and 0.50 at L$^\prime$ band and K$^\prime$ band, respectively. These fractional fluxes differ significantly from those estimated by the geometric models. This is not a physical feature of V892 Tau, but rather a bias in the SQUEEZE algorithm that we quantify using simulations (Sections \ref{sec:imrecons} and \ref{subsec:squeezemod}).}\label{fig:squeeze}
\end{figure*}

\begin{figure*}[ht!]
\plottwo{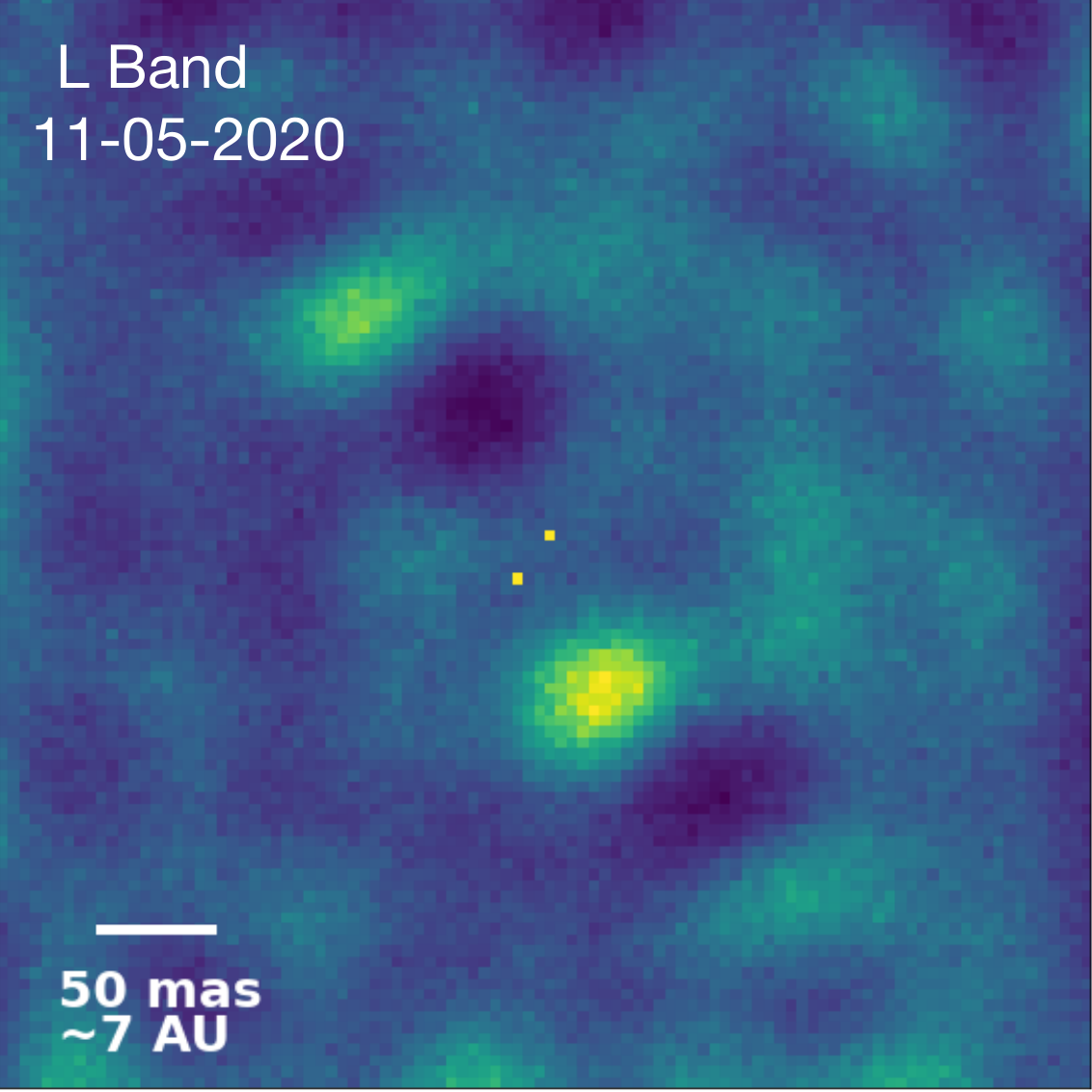}{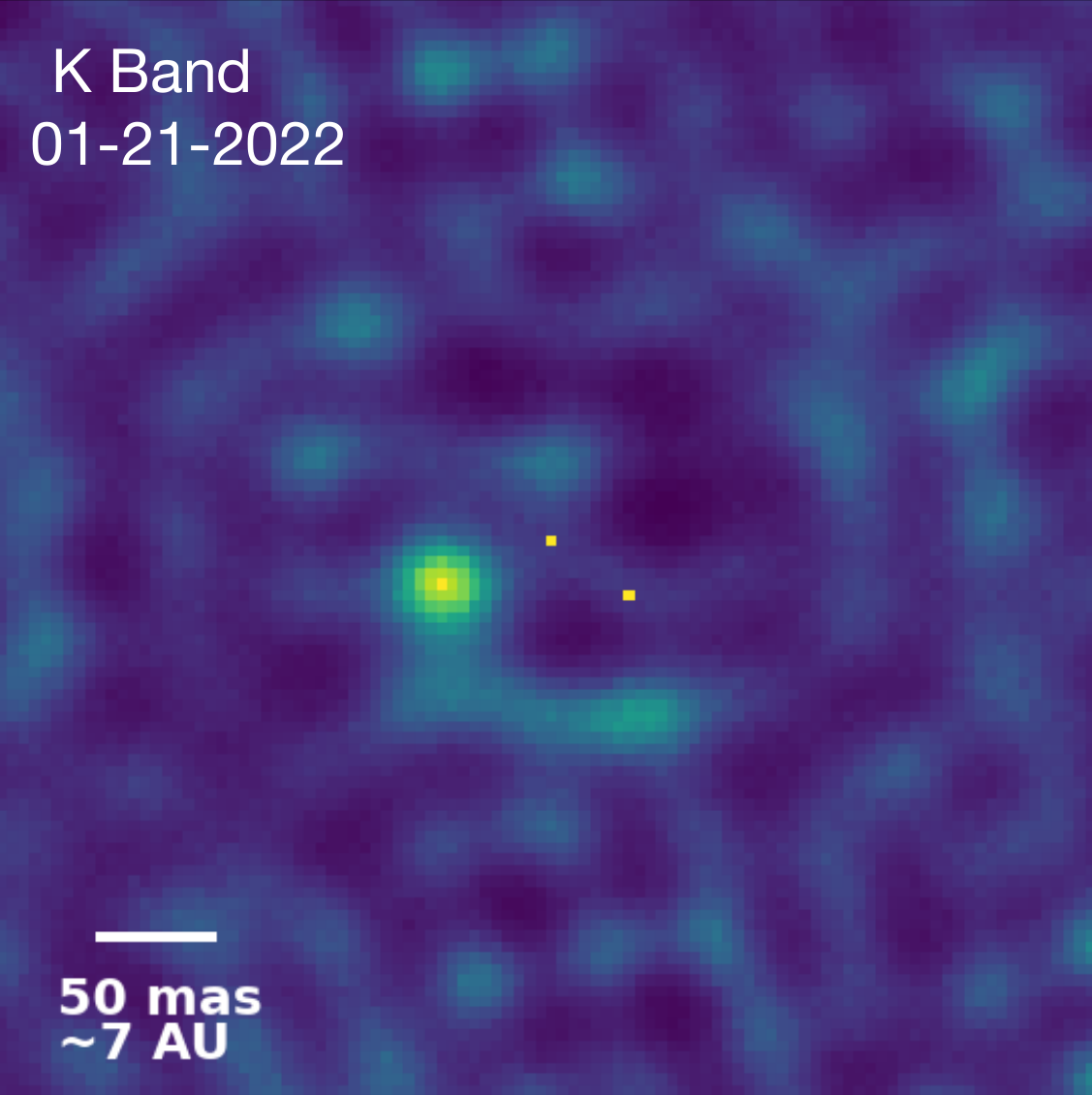}
\caption{SQUEEZE images reconstructed with a binary model from the L$^\prime$ band (left) and K$^\prime$ band (right) observables. The central stars and companions are represented with $\delta$ functions equal to their fractional fluxes and the images are normalized appropriately.  Changes in the separations and position angles of the $\delta$ functions show the orbital motion of the stellar companion between the two epochs. The fractional fluxes of the central stars are  0.51 and 0.62 at L$^\prime$ band and K$^\prime$ band, respectively. The fractional fluxes of the secondary components are 0.28 and 0.25 at L$^\prime$ band and K$^\prime$ band, respectively.}\label{fig:squeezebinary}
\end{figure*}

\begin{figure*}[ht!]
\plotone{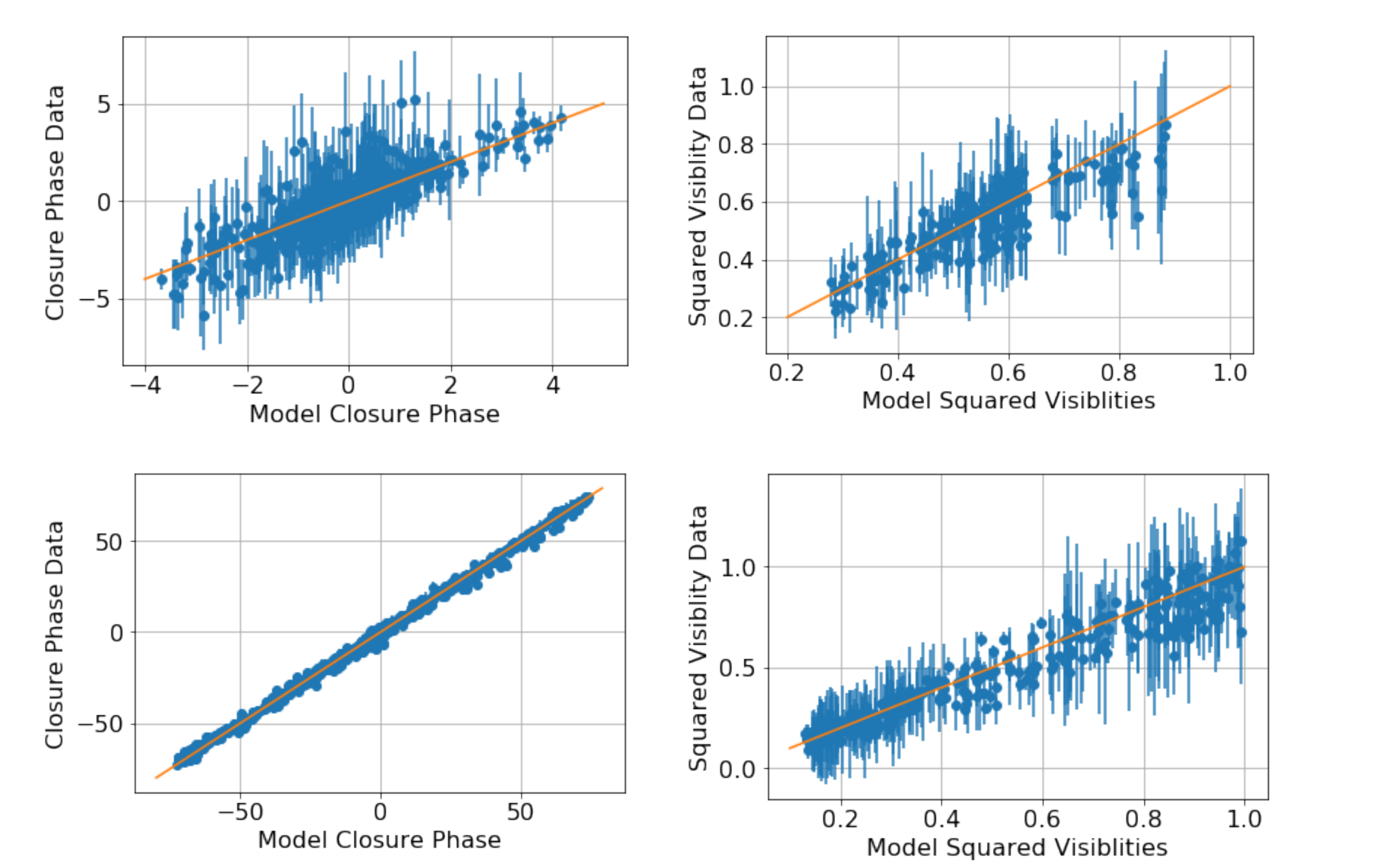}
\caption{Top left: Scattered points with error bars in the left panels show the correlation between the closure phase data and the best-fit model closure phases at L$'$ (top) and K$'$ (bottom).  Scattered points with error bars in the right panels show the correlation between the squared visibility data and the best-fit model squared visibilities at L$'$ (top) and K$'$ (bottom).  In all panels the orange solid line represents a 1:1 correlation.}\label{fig:correlation}
\end{figure*}
%\begin{figure*}[ht!]
%\plottwo{Lband_geocropped.png}{kband_geomodelcropped.png}
%\caption{}\label{fig:squeeze}
%\end{figure*}

\section{Results}\label{sec:results}

\subsection{Fourier Observables and Reconstructed Images}\label{sec:imrecons}

Figure \ref{fig:results} shows the Fourier observables - closure phases (left) and squared visibilities (right) - at both wavelengths. At L$^\prime$ band, the squared visibilities fall off rapidly as a function of baseline length independent of position angle, which is characteristic of a centro-symmetric, extended morphology. At K$^\prime$ band, some baselines fall off more sharply than others, indicating a lower degree of centro-symmetry. The closure phases at both bands have high values compared to the calibrators, which suggests asymmetry at close-in angular separations.   %In the following subsections, we discuss the best-fit models of the V892 Tau morphology, and their ability to reproduce the observables and the reconstructed images.

 Figure \ref{fig:squeeze} and Figure \ref{fig:squeezebinary} show the reconstructed images that include the SQUEEZE single point source model and binary model, respectively. These images also suggest that the K$^\prime$ band morphology is less centro-symmetric than that at L$^\prime$ band. Approximately $\sim36\%$ of the non-stellar flux is enclosed in the compact signal located southwest of the star at K$^\prime$ band, compared to  $\sim6.4\%$ in the L$^\prime$ band signal located southeast of the star. In Figure \ref{fig:squeezebinary}, the compact signals are removed from the images since they are captured by the SQUEEZE binary model. The K$^\prime$ image has a higher fraction of the total flux accounted for by the binary components, at 0.87 compared to 0.79 at L$^\prime$.
 %total fractional fluxes of the central star plus companion are 0.79 and 0.87 at L$^\prime$ band and K$^\prime$, respectively. 
 The residual extended emission in the L$^\prime$ band image and the removal of the southwest compact signal in the K$^\prime$ images are consistent with the single point source model images. 

The SQUEEZE images show different morphologies between the datasets and the models used during reconstruction (single point source versus binary). The binary model places the companion at 26.9 $\pm$ 0.7 mas with a PA of 144.9 $\pm$ 2$^{\circ}$ at L$^\prime$ band and 39.6 $\pm$ 0.03 mas with a PA 239.8 $\pm$ 1$^{\circ}$ at K$^\prime$ band. The changes in the separation and position angle of the companion are due to orbital motion between the two epochs. 
We find agreement between the positions of the companion delta functions in the binary model images and the locations of the compact signals in the single point source model images, but the fractional fluxes of the central star differ between the two models. With the single point source model the fractional fluxes of the central stars are 0.71 and 0.50 at L$^\prime$ and K$^\prime$ band, respectively. Using the binary model, the fractional fluxes of the central stars are  0.51 and 0.62, with secondary fractional fluxes of 0.28 and 0.25, at L$^\prime$ and K$^\prime$ band, respectively. 

%due to biases introduced in the fitting process (discussed in the following paragraph and Section \ref{subsec:squeezemod}). 

As we further demonstrate in Section \ref{subsec:squeezemod}, these discrepancies are due to biases introducted in the fitting process. SQUEEZE has difficulty simultaneously matching the squared visibilities and closure phases when the single point source model is used ($\chi^{2}$$=$559.99). The binary model provides a better match to the data ($\chi^{2}$$=$327.85) and is also a more appropriate model given the known stellar companion. In the single point source model, achieving a better match to both observables would require arbitrary changes to the error bar scalings of the squaredvisibilities and closure phases \citep[e.g.][]{2017ApJS..233....9S}. Since the error bar scaling process would be motivated not by the data but by the SQUEEZE algorithm, we refrain from doing this and instead show all reconstructions using original error bars.

 Although the use of the binary model is more physically motivated, we include both SQUEEZE models to demonstrate the effects of adding different components during the reconstruction process. While the single point source model allows us to more freely place circumstellar emission at any location in the images, the binary model images reveal complex structure in the disk that is not visually apparent in the single point source images (Figure \ref{fig:squeezebinary}). The K$^\prime$ band image shows a point-like feature to the southeast of the central star, and the L$^\prime$ band image shows complex asymmetries in the form of multiple arcs. 
 To explore these features in the context of different disk plus companion scenarios, we perform simulated image reconstructions described in Section \ref{subsec:squeezemod}.
 
 %that, as discussed in Section 4.3, is consistent with a circumprimary disk, and the L$^\prime$ band image shows complex structure that cannot be easily explained by a smooth, geometric circumbinary disk model. To ensure that the geometric models are indeed consistent with the SQUEEZE images, we perform simulated image reconstructions described in Section \ref{subsec:squeezemod}.

%We compare the fractional fluxes in Figure \ref{fig:squeeze} to those in Figure \ref{fig:squeezebinary}. From the binary model, the fractional fluxes of the central stars are 0.31 and 0.49 at L$^\prime$ and K$^\prime$ band, respectively. The fractional fluxes of the companion are 0.49 and 0.23 at L$^\prime$ and K$^\prime$ band, respectively. At L$^\prime$, bright emission from the disk is causing the companion to have a higher fractional flux than the central star. The binary model places the companion at 26.9 mas with a PA of 144.9$^{\circ}$ and 39.6 mas with a PA 239.8$^{\circ}$ at L$^\prime$ and K$^\prime$ band, respectively. We find that both models are consistent with each other, however the binary model better resolves disk structure.

\subsection{Best-fit Geometric Models} \label{sec:bestfit}
Table \ref{table:3} lists the best-fit parameters and goodness-of-fit metrics for each of the three model scenarios (companion-only, disk-only, and disk-plus-companion).
Examining the $\chi^{2}$ values in Table \ref{table:3}, the squared visibilities and closure phases at both wavelengths are best described by the disk-plus-companion model. To assess the significance of the preference for the disk-plus-companion model over the others, we compare the improvements in $\chi^2$ values between models to distributions with N degrees of freedom, where N is the difference in number of parameters between two models. We find that the disk-plus-companion model is preferred at $>$5$\sigma$ for both wavelengths.

At K$^\prime$ band, we find that the reduced $\chi^{2}$ values are closest to 1 for the disk-plus-companion model, ranging between 1 and 2 for the two Fourier observables.
At L$^\prime$ band, the individual disk-plus-companion reduced $\chi^{2}$ values for the closure phases and squared visibilities are $<1$, which would imply over-fitting for perfect error bars. However, the conservative error bar calculations applied here (see Section \ref{sec:datared}) may bias the reduced $\chi^{2}$ toward low values. 
We thus base the $\chi^2$ model selection primarily on the raw, rather than reduced values.
Figure \ref{fig:correlation} shows the correlation between the disk-plus-companion model and the data for both observables and bands.
Given the conservative error bars, the $\chi^2$ model selection tests, and these correlations, we  accept the disk-plus-companion model as the best fit for both wavelengths.

\begin{figure*}[ht!]
\plottwo{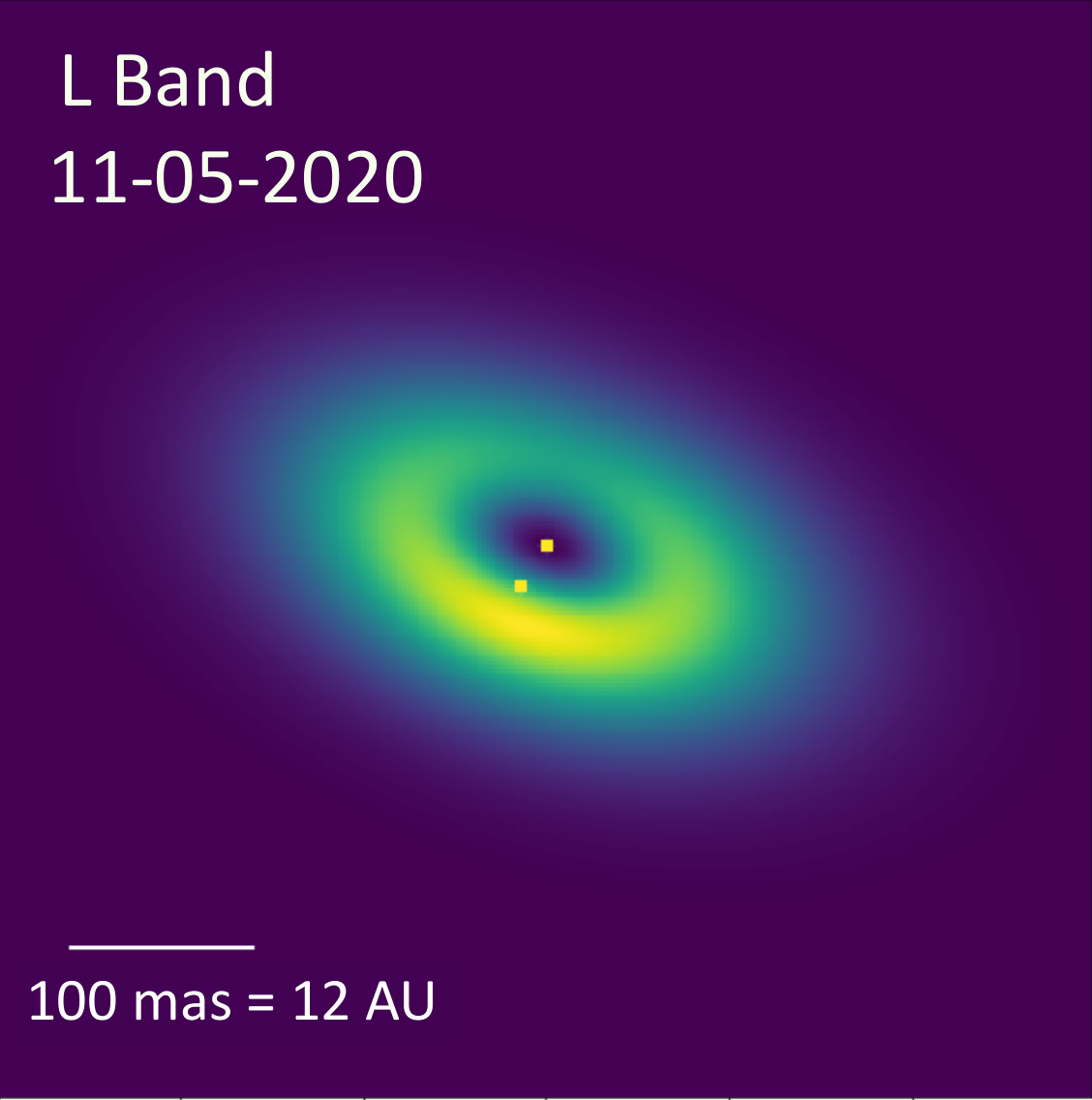}{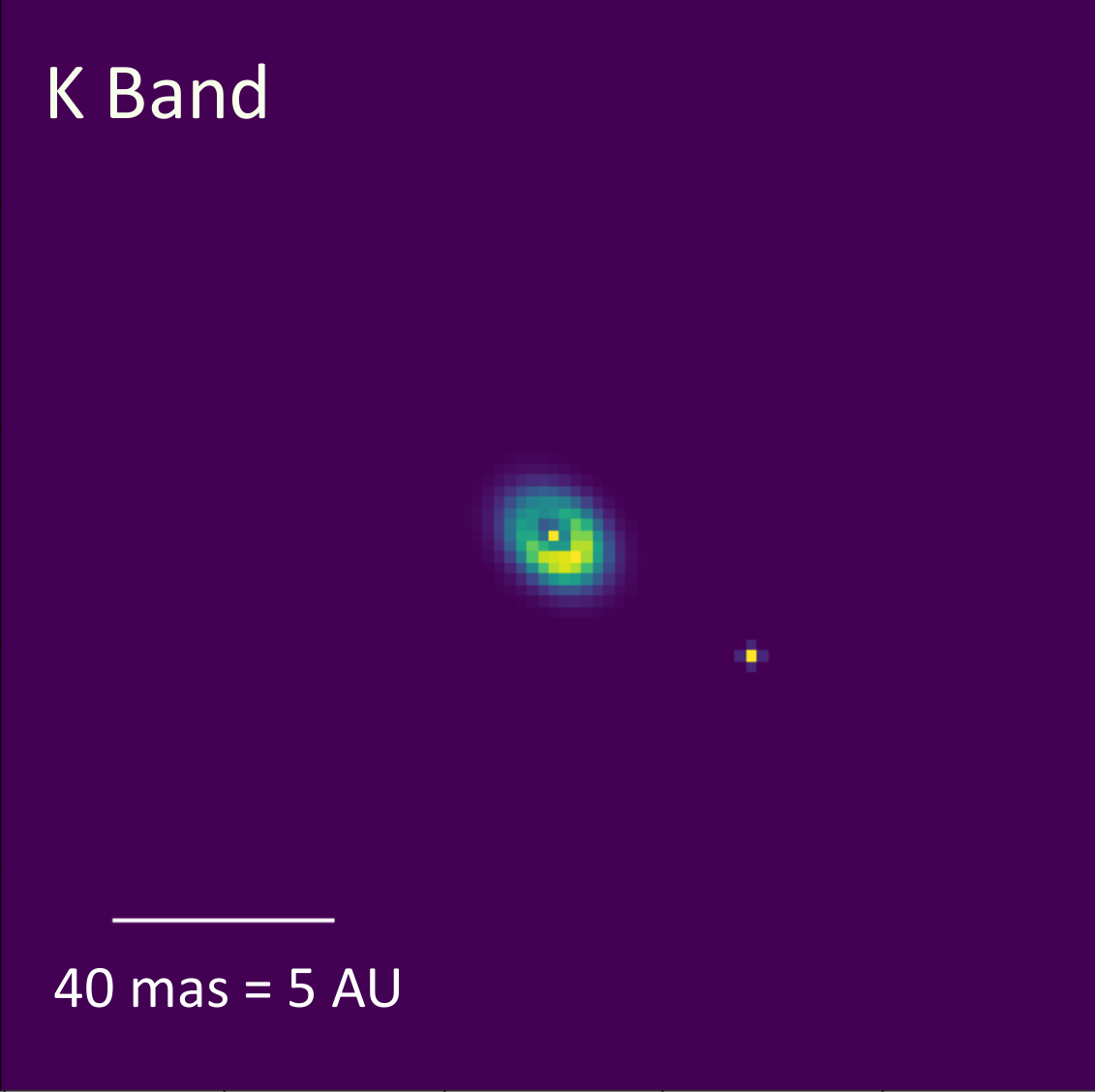}
\caption{Left: The best fit geometric model at L$^\prime$ band. The FWHM of the circumbinary disk is  189.9 +16.5/-19.7 mas and the separation of the stellar companion is 26.84 +0.75/-0.64 mas at 146.6 +1.35/-1.32 degrees measured east-of-north. The image has been sub-framed and smoothed with a Gaussian to make the $\delta$ functions visible. Right: The best fit geometric model at K$^\prime$ band. The circumprimary disk has FWHM of  15.7 +2.3/-2.0 mas, and a separation of  42.1 +0.70/-0.63 mas located at 238.51 +0.98/-0.83 degrees measured east-of-north. The image has been sub-framed and smoothed with a Gaussian to make the $\delta$ functions visible. 
Changes in the separations and position angles of the companions between the two epochs show the orbital motion of the stellar companion.
The locations of the companion in both panels are similar to those in Figure \ref{fig:squeezebinary}; however their appearances are stretched here due to differences in the images' fields of view.}\label{fig:geo}
\end{figure*}

Figure \ref{fig:geo} shows the best fit models at L$^\prime$ band and K$^\prime$ band. At L$^\prime$ band, the geometry of the system is best described by a circumbinary disk and companion. We find that the FWHM of the CB disk is 189.9 +16.5/-19.7 mas ($\sim$ 25.5 AU) and the separation of the stellar companion at the time of observation is  26.0 +0.7/-0.6 mas located at 147.4$^{\circ}$ $\pm$ 1.4$^{\circ}$ measured east of north. The contrast of the stellar companion is 0.60 $\pm$0.03  magnitudes relative to the host star. From the total flux of the system and the contrast, the flux of the secondary companion is calculated and converted to an absolute magnitude, M$_{L}$ = 6.63 $\pm$0.03  mag.

At K$^\prime$ band, we find that the geometry of system is best represented with a circumprimary disk and companion. We detect a circumprimary disk with a FWHM of 15.7 +2.3/-2.0 mas ($\sim$ 2 AU), and a companion with a separation of  42.1 +0.70/-0.63 mas ($\sim$ 5.6 AU) located at 238.51$^{\circ}$ +0.98/-0.83$^{\circ}$ east of north. The contrast of the secondary star relative to the primary star is 0.67 +0.02/-0.06 mag, giving M$_{K}$ = 6.49 +0.02/-0.06  mag. The skew of the circumprimary disk is roughly aligned with the PA of the companion.

To test that the detection of the circumprimary disk is a physical feature of V892 Tau and rule out a local minimum in the fitting, we explore a set of models where we set the upper bound of the companion separation (S) prior to:
\begin{equation}
    S < \mathrm{FWHM} \frac{a_{ratio}*f_{h}}{2\sqrt{a_{ratio}^{2}\cos^{2}\alpha+\sin^{2}\alpha}}.
\end{equation}
where
\begin{equation}
    \alpha=\mathrm{PA}-\theta.
\end{equation}
In the above equations, $f_{h}$ is the fraction of the semi-major axis that is occupied by a hole and $\theta$ is the position angle of the major axis of the disk. These equations ensure that companion is always within the disk, forcing a CB disk. 

Table \ref{table:3} lists the results of the forced circumbinary disk-plus-companion model fit to the K$^\prime$ band data.  The $\chi^{2}$ values indicate that this model does not adequately describe the data. The discrepancy between the data and model is especially apparent in the closure phases, where the reduced $\chi^{2}$ value, $\chi^{2}_{r}$, is 5.11 for the forced CB model, as opposed to 2.18 for the unrestricted disk-plus-companion model. The $\chi^{2}$ difference between the unrestricted disk-plus-companion model and the forced CB disk model is 2240.16. We compare this value to significance estimates for a $\chi^{2}$ distribution with 10 degrees of freedom, the number of parameters in the CB disk-plus-companion model. The circumprimary disk scenario is preferred at $>$5$\sigma$.  Furthermore, the forced circumbinary parameters are poorly constrained due to the existence of many local likelihood maxima, which also include non-physical scenarios.
We thus find the circumprimary disk detection to be robust.

\begin{deluxetable*}{ccccc}
 \tablecaption{}
\tablehead{ \colhead{Parameter} & \colhead{Companion}& \colhead{Disk}& \colhead{Companion+Disk}&\colhead{Companion+Disk (forced CB)}}
\startdata
\multicolumn{5}{c}{L Band Fit Results}\\
 \hline
S (mas) & 52.1 $\pm$ 3 & - &  26.0 +0.7/-0.6 & -\\
PA ($^{\circ}$) &157.3 +2.5/-1.6 &-&147.4 $\pm$ 1.4 & -\\
CC (mag) & 4.1 $\pm$ 0.1 &-& 0.60 $\pm$0.03 & -\\
FWHM (mas) & - & 75.3 +4.2/-2.9 & 189.9 +16.5/-19.7 & -\\ 
a$_{ratio}$ & -& 0.55 $\pm$ 0.02 & 0.67 $\pm$ 0.09 & -\\
$\theta$ ($^{\circ}$)&-& 137.6 +1.6/-1.9&  74.8 +8.1/-5.8 & -\\
b&-&0.64 $\pm$ 0.01& 0.50 $\pm$ 0.007 & -\\
A$_{s}$&-&  0.27 $\pm$ 0.02 & 0.24 $\pm$ 0.05 & -\\
$\phi_{s}$ ($^{\circ}$)&-&170.2 +3.2/-4.5& 169.5 +4.1/-4.0&-\\
f$_{h}$&-&0.89 $\pm$ 0.08& 0.48+ 0.1/-0.07&-\\
$\chi^{2}$& 4882.30&797.94&423.66 &-\\
DOF&717&713&710&-\\
CP $\chi^{2}$&812.85 &467.63 & 336.47&-\\
CP $\chi_{r}^{2}$ &1.62 & 0.94& 0.68& -\\
V$^{2}$ $\chi^{2}$& 4069.44& 330.30 & 87.19& -\\
V$^{2}$ $\chi_{r}^{2}$ &19.10 &1.58 & 0.42&-\\
%Evidence Values&-336.9 &-1261.32 & -819.38&-\\
\hline
\hline
\multicolumn{5}{c}{K Band Fit Results}\\
\hline
S (mas) & 42.0 $\pm$0.6 &- & 42.1 +0.70/-0.63 & 40.5 +0.9/-38.0 \\
PA ($^{\circ}$) &237.83 +0.87/-0.81 &-& 238.51 +0.98/-0.83 &237.91 +3.91/-196.43 \\
CC (mag) & 1.064 $\pm$ 0.04&-&0.67 + 0.02/-0.06 & 0.41 +0.56/-0.0039\\
FWHM (mas) &-&64.2 +0.5/-0.3& 15.7 +2.3/-2.0 &  95.98 +55.67/-29.29\\ 
a$_{ratio}$ & -&0.123 $\pm$ 0.004& 0.70 +0.05/-0.03 & 0.83 +0.15/-0.80 \\
$\theta$ ($^{\circ}$)&-&59.58 $\pm$ 0.04& 53.35 +3.2/-2.5& 59.46 +6.53/-11.06 \\
b&-&0.000198 $\pm$ 0.0013&0.41 $\pm$ 0.04 & 0.64 +0.007/-0.48 \\
A$_{s}$&-&0.29 +0.01/-0.04&0.38 $\pm$ 0.04& 0.92 +0.06/-0.16   \\
$\phi_{s}$ ($^{\circ}$)&-&198.8 +2.9/-0.7&196.3 +5.8/-5.4 &155.9 +43.10/-109.48  \\
f$_{h}$& -&0.34 +0.007/-0.02&0.68 +0.2/-0.3 &0.97 +0.01/-0.26\\
$\chi^{2}$&23830.28&44112.85& 2026.75&4266.91\\
DOF &1077&1073&1070&1070\\
CP $\chi^{2}$& 16189.34&40475.99&1626.61 &3815.82\\
CP $\chi_{r}^{2}$ &21.49 &54.04 & 2.18&5.11 \\
V$^{2}$ $\chi^{2}$&7640.94 &3636.85 &400.14& 451.09\\
V$^{2}$ $\chi_{r}^{2}$ &23.80 &11.32 &1.27 & 1.43\\
%Evidence Values&-356.5 &-47290.85 & -20070.66&-52085.515\\
\hline
\hline
\enddata
\label{table:3}
\caption{The maximum likelihood models for each scenario. Degrees of freedom is indicated by DOF. Closure phases and squared visibilites are denoted by CP and V$^{2}$, respectively. Chi-squared and reduced chi-squared values are indicated by $\chi^2$ and $\chi^2_r$, respectively.}
\end{deluxetable*}

\begin{figure*}[ht!]
\plottwo{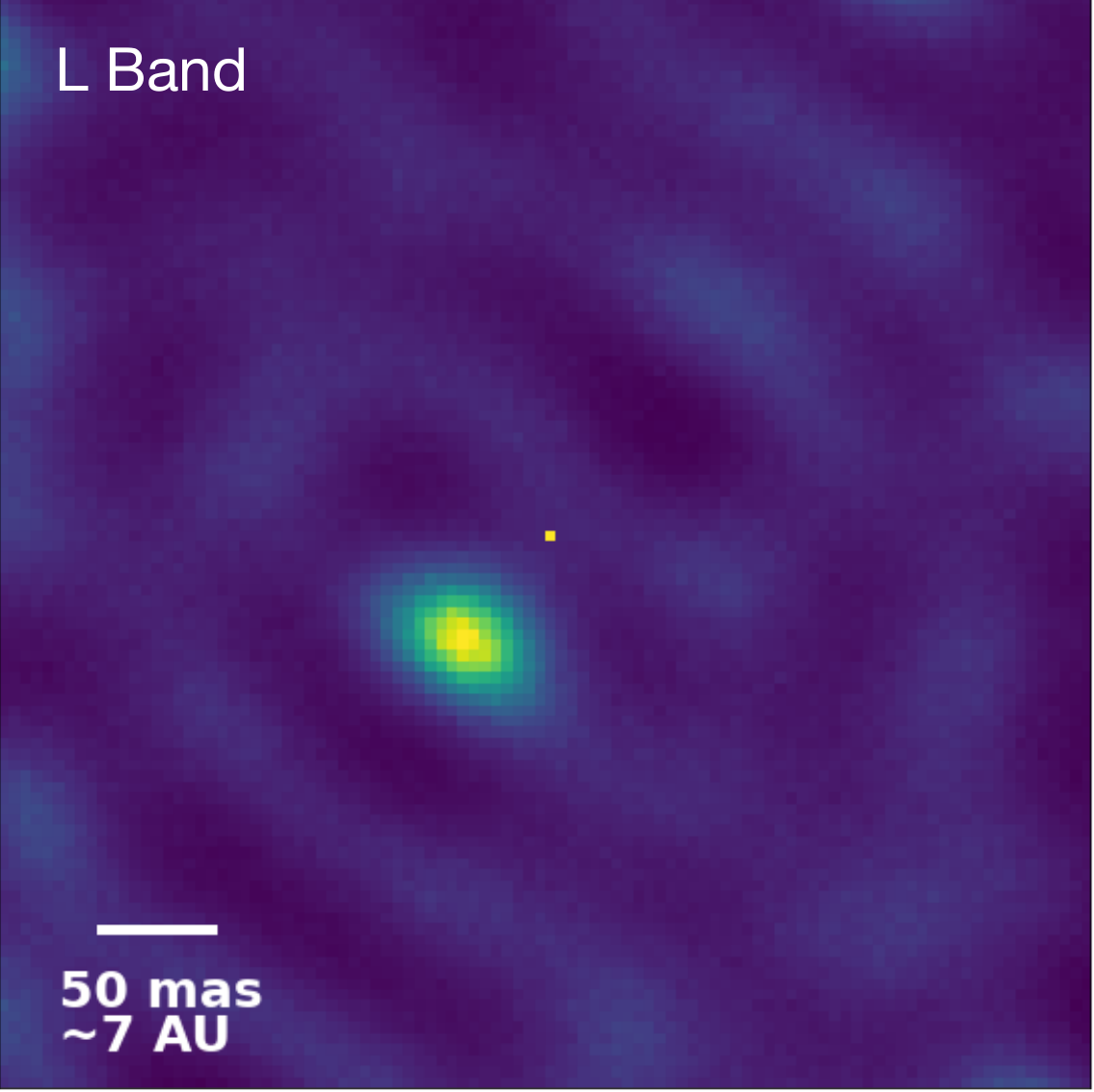}{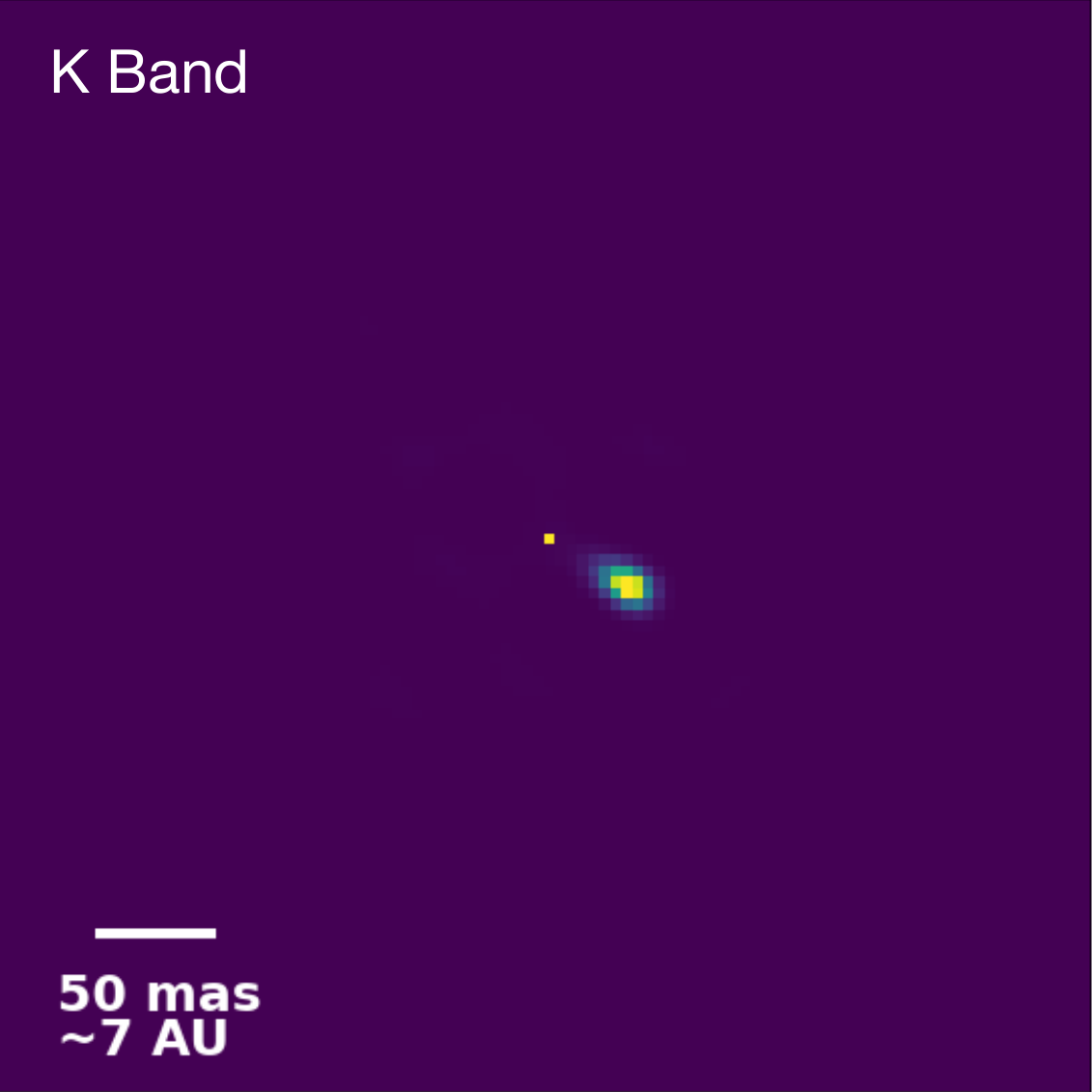}
\caption{SQUEEZE images reconstructed from the disk-plus-companion model closure phases and squared visibilities at L$^\prime$ band (left) and K$^\prime$ band (right) using a single point source model. The fractional fluxes of the central stars are 0.79 at L$^\prime$ band and 0.51 at K$^\prime$ band. This approximately matches the fractional fluxes in the SQUEEZE reconstructions of the data.}\label{fig:modsqueeze}
\end{figure*}

\begin{figure*}[ht!]
\plottwo{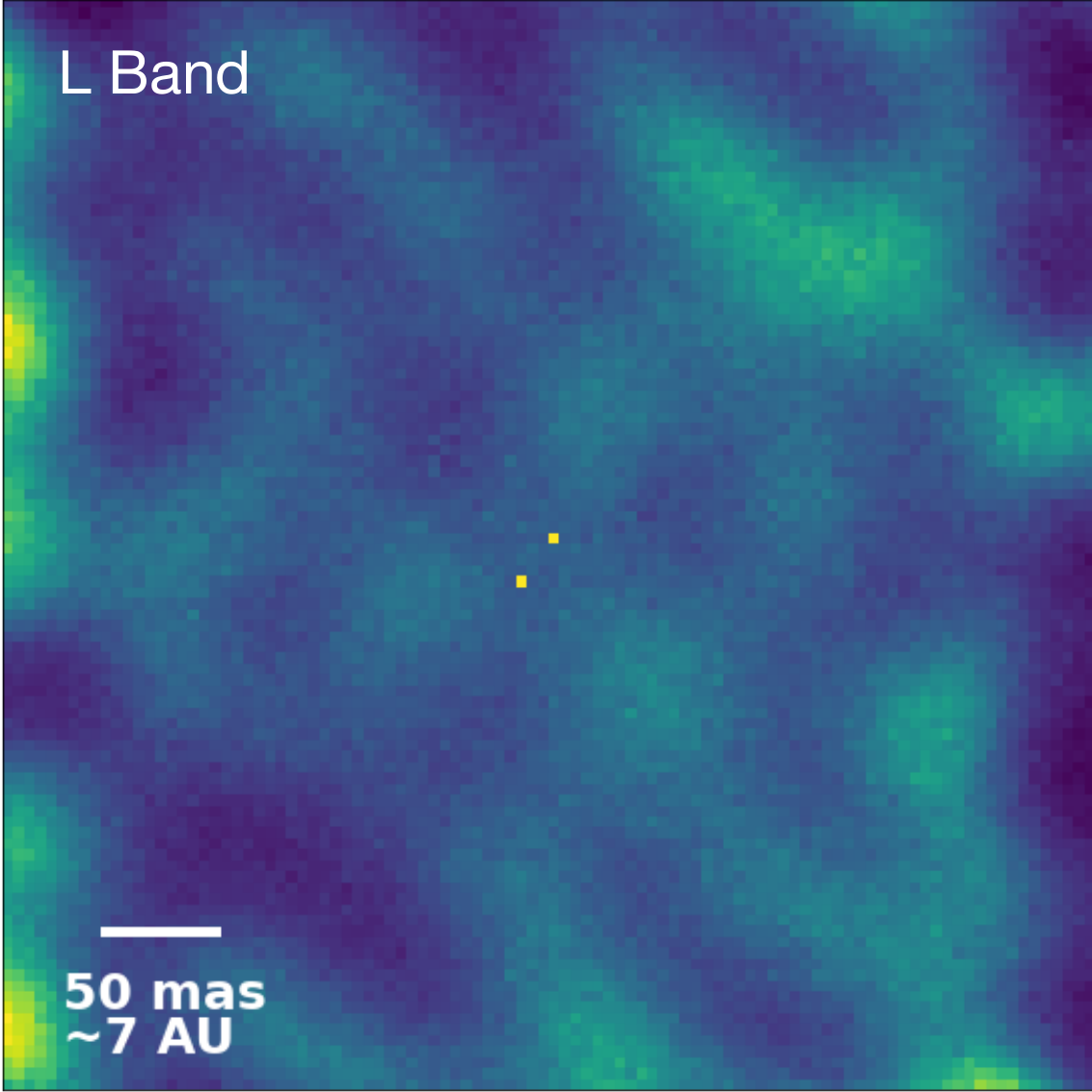}{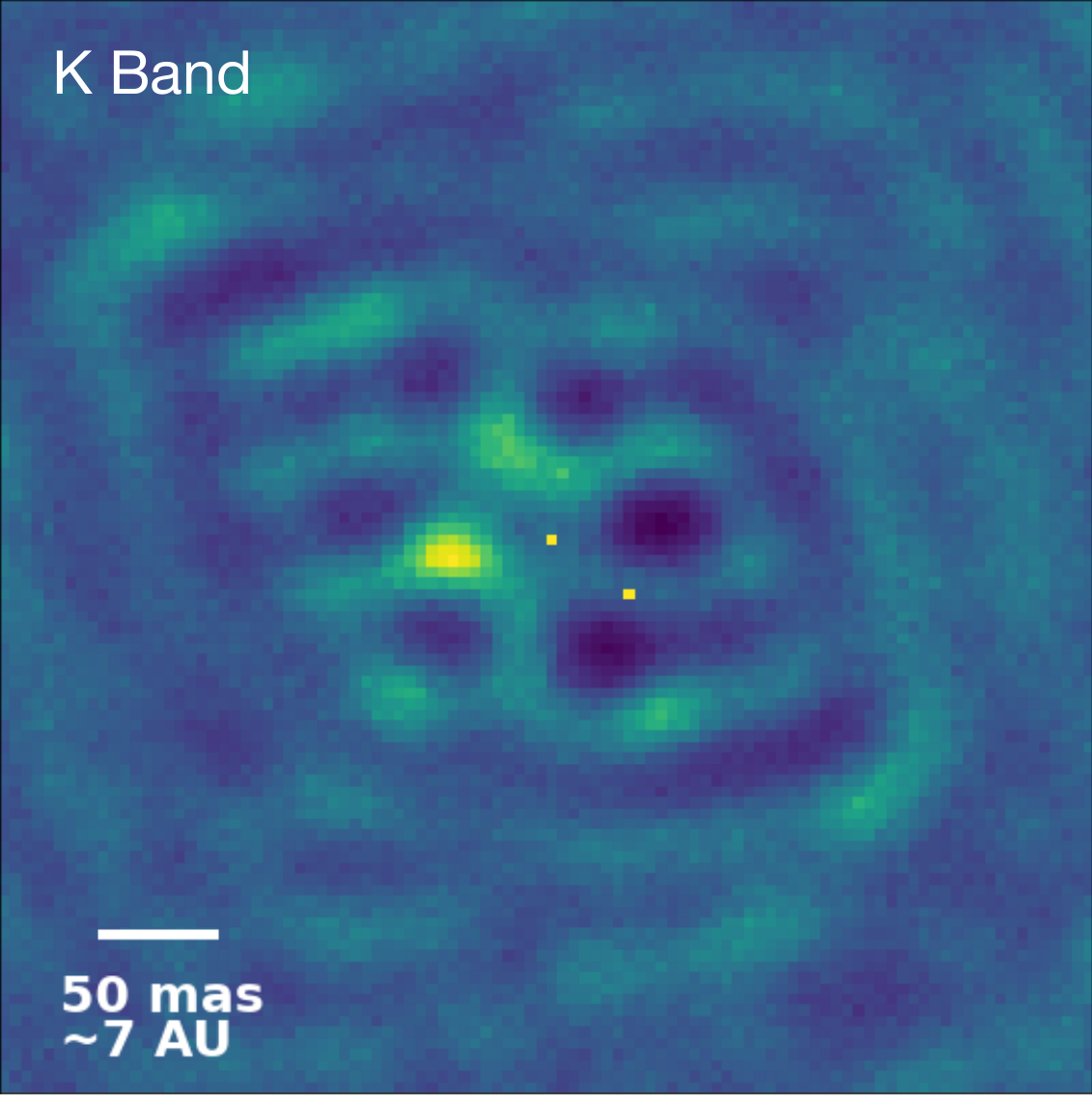}
\caption{SQUEEZE images reconstructed from the disk-plus-companion model closure phases and squared visibilities at L$^\prime$ band (left) and K$^\prime$ band (right) using a binary model. The fractional fluxes of the central stars are  0.57 at L$^\prime$ band and 0.65 at K$^\prime$ band. The fractional fluxes of the companion are 0.31 at L$^\prime$ band and 0.27 at K$^\prime$ band. This approximately matches the fractional fluxes in the SQUEEZE reconstructions of the data.}\label{fig:modsqueezebinary}
\end{figure*}

\begin{figure*}[ht!]
\plotone{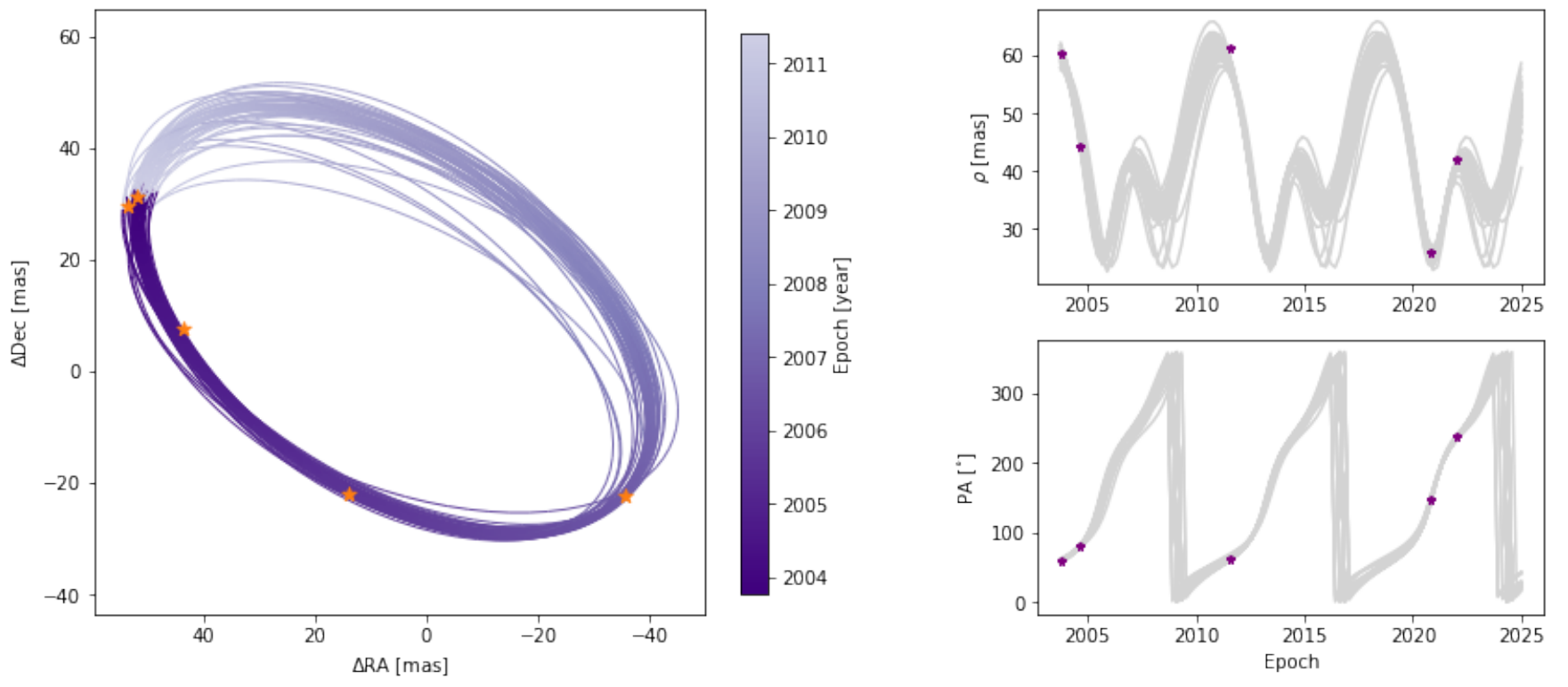}
\caption{Left: 50 random orbits sampled from the distribution using Orbitize!. The orange stars represent all astrometric measurements of V892 Tau to date. The color bar represents one orbital period ($\sim$7 years). Top Right: The multi-epoch separation of the stellar companion and predictions of future locations. Bottom Right: same as top right but for position angle. We find that the orbit is consistent with the astrometry estimated by the SQUEEZE images as well. }\label{fig:orbit}
\end{figure*}

\begin{deluxetable}{lcc}
\caption{V892 Tau astrometry, showing archival constraints and the L$^\prime$ and K$^\prime$ band astrometry presented here.}
\tablehead{
\colhead{Reference} & \colhead{Separation ($"$)} & \colhead{PA ($^\circ$; East-of-North)}} 
\startdata
Smith et al. 2005$^{*}$ & 50.5$\pm$ 4&234$\pm$ 3  \\
Smith et al. 2005 & 60.4 $\pm$ 1 & 59 $\pm$ 1  \\
Monnier et al. 2008 &44.2 $\pm$ 1 &79.9 $\pm$ 1\\
Long et al. 2020 & 61.3 $\pm$ 3 & 61 $\pm$ 3\\
This work (L$^\prime$  band) & 26.8 $\pm$ 0.7 & 146.6 $\pm$ 1.3\\
This work (K$^\prime$ band) & 41.3 $\pm$ 0.7 & 239.03 $\pm$ 0.83
\label{table:astrometry}
\enddata
\footnotesize{$^*$ We omit the first row from orbital fitting due to a $180^\circ$ ambiguity in the position angle.}
\end{deluxetable}  

\begin{figure*}[ht!]
\plottwo{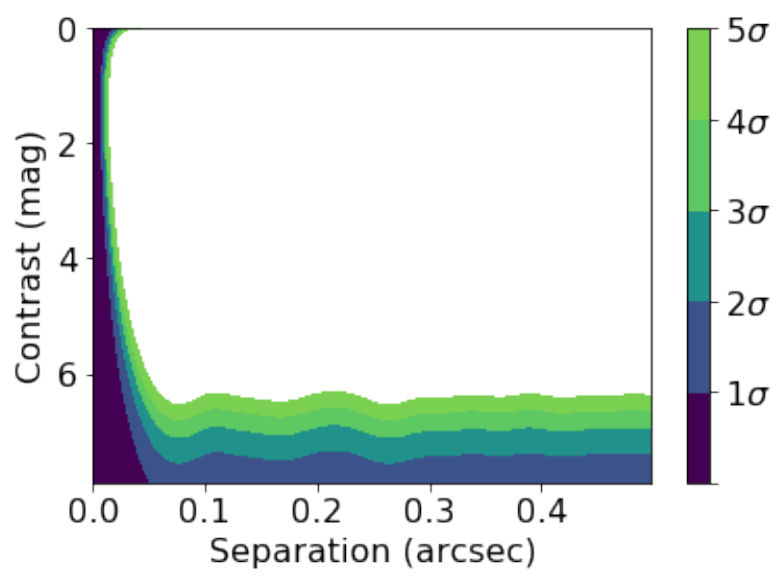}{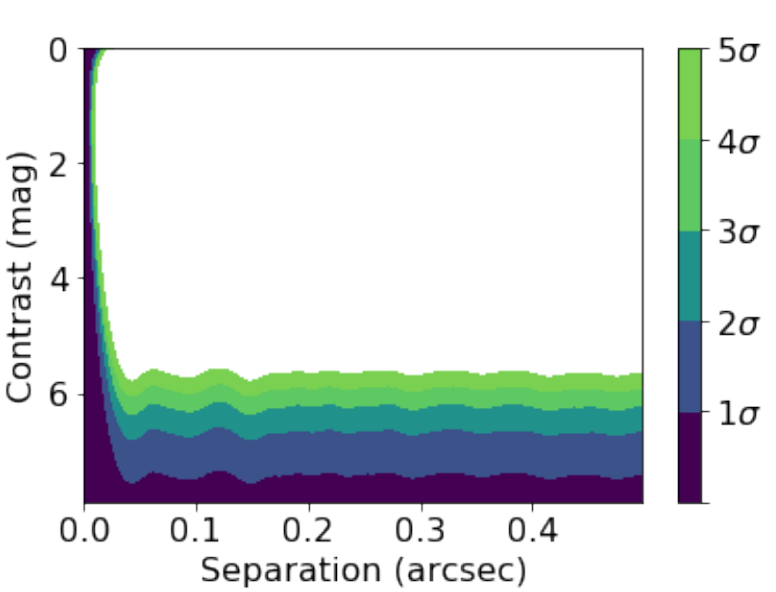}
\caption{Contrast curves for V892 Tau at L$^\prime$  band (left) and K$^\prime$ band (right) using NRM and the PyWFS. We find that the PyWFS contrast is 0.5-1 magnitudes better than observations with NIRC2 NRM behind the Shack-Hartmann wavefront sensor. We use these contrast curves to place limits on undetected companions and find that we are sensitive to $\gtrsim$20 M$_{J}$ planets at L$^\prime$ band and $\gtrsim$50 M$_J$ planets at K$^\prime$ band (assuming an age of $\sim1$ Myr).}\label{fig:cc}
\end{figure*}
%\begin{figure*}[ht!]
%\plotone{epoch.png}
%\caption{}\label{fig:results}
%\end{figure*}

\subsection{SQUEEZE Model Reconstruction}\label{subsec:squeezemod}
Image reconstructions capture the true source brightness distribution to varying degrees depending on the mask (u,v) coverage, the amount of sky rotation in the observations, and the particular regularization choices in the individual algorithms \citep[e.g.][]{2017ApJS..233....9S}.
We thus check whether the best-fit geometric model reproduces the structure in the images reconstructed from the observations (Figure \ref{fig:squeeze} and Figure \ref{fig:squeezebinary}). We generate model closure phases and squared visibilities by sampling the best fit geometric models with the same sky rotation and (u,v) coverage as the observations. We then add noise to the model closure phases and squared visibilities such that the scatter matches that of the observations. Figures \ref{fig:modsqueeze} and \ref{fig:modsqueezebinary} show the resulting images, which are generally consistent with the reconstructions generated from the data (Figure \ref{fig:squeeze} and \ref{fig:squeezebinary}).

The fractional fluxes of the central stars in the simulated reconstructions using SQUEEZE's single point source model (0.79 at L$^\prime$ band and 0.51 at K$^\prime$ band) are similar to those for the data (0.71 at L$^\prime$ band and 0.50 at K$^\prime$ band). In the reconstruction using SQUEEZE's binary model, the fractional fluxes are also similar to the observations, with central star fractional fluxes of 0.57 at L$^\prime$ band and 0.65 at K$^\prime$ band. The fractional fluxes of the companions are 0.31 at L$^\prime$ band and 0.27 at K$^\prime$ band. 
This demonstrates in a controlled way that the variation in fractional fluxes for different SQUEEZE configurations is an algorithmic bias. Furthermore, the consistency between the fractional fluxes for the simulations and the observations shows that approximately the same algorithmic biases are introduced in the reconstructions generated from the data and from the geometric models.

We find that when we reconstruct the image from the best-fit disk plus companion observables using a single point source SQUEEZE model, the model reconstruction visually matches the data at both bandpasses. However, this is not the case at both bands when we reconstruct images for the disk plus companion model observables using SQUEEZE's binary model. At K$^{\prime}$ band, we find that the best-fit circumprimary disk model reproduces the data, including the point-like blob to the southeast of the star. We rigorously test this by reconstructing the image from the best-fit binary geometric model, and find that the point-like blob in Figure \ref{fig:modsqueezebinary} is removed from the image. At L$'$ band, the simulated reconstruction lacks the complex structure discussed in Section \ref{sec:imrecons}, indicating that it cannot be captured by the geometric model. We provide our interpretation of this in Section \ref{subsec:systemgeometry}.

\subsection{Orbit Fitting}
  We update the orbit of the V892 Tau stellar companion using \texttt{Orbitize!}~\citep[e.g.][]{2020AJ....159...89B}. We fit our astrometry and the archival data shown in Table \ref{table:astrometry}, which lists the separations and position angles, measured east of north, of each data point included in the orbit fit. We exclude the first data point in \cite{2005A&A...431..307S}, following a similar method as \cite{2021ApJ...915..131L}. This data point was ambiguously measured; it was unclear which component was the primary or secondary star, making the uncertainty in  the position angle 180$^{\circ}$.
  
We use parallel-tempered MCMC methods to fit an orbit to the data, with 10 temperatures, 100 walkers, and 10,000 steps. We use the \texttt{Orbitize!}~default uniform priors; varying the semi-major axis from 0.001 to 10$^{7}$ AU, the eccentricity from 0 to 1, the inclination from 0 to 2$\pi$, the argument of periastron from 0 to 2$\pi$, the position angle of the ascending node from 0 to 2$\pi$ (measured E of N), and the periastron passage from 0 to 1. The distance and total mass of the system are Gaussian priors that are centered on the parallax \citep[7.4 $\pm$ 0.08 mas;][]{2020yCat.1350....0G} and total mass of the system \citep[6.0 $\pm$ 0.2 \(M_\odot\);][]{2021ApJ...915..131L}, respectively. Parallax uncertainties for binaries with separations less than a few arcseconds and G $\lesssim$ 18 have been underestimated by $\lesssim$ 30$\%$ \citep[][]{2021MNRAS.506.2269E}.  Thus, the parallax error is likely slightly underestimated due to the presence of the binary and disk. However, since the astrometric error bars dominate the orbital error budget, even a 30$\%$ inflation of the parallax error would not significantly change the orbit fit results.

The left panel of Figure \ref{fig:orbit} shows 50 random orbits in the posterior distribution with the color bar representing a single orbit. The right panel shows the separation and position angle of V892 Tau's stellar companion across all the epochs. 
The semi-major axis of the stellar companion is 6.8 +0.04/-0.03 AU (49.0 $\pm$ 2 mas) with a period of 7.2 +0.07/-0.05 years. The eccentricity of the orbit is 0.26 $\pm$0.04 and its inclination is 58.4$^{\circ}$ $\pm$ 3$^{\circ}$. From these estimates, we independently measure the dynamical mass of the system to be 6.1 +0.2/-0.1  \(M_\odot\), which is in agreement with measurements made by \cite{2021ApJ...915..131L}. We compare our values to previous estimates in Table \ref{table:orbitresults}.

\subsection{Additional Companions}
 Figure \ref{fig:cc} shows the 1-5$\sigma$ contrast curves at L$^\prime$ band (left) and K$^\prime$ band (right). We convert the contrast limits to absolute magnitudes and compare them to magnitudes predicted by hot start models with an assumed age of 1 Myr \citep[approximately the same age as V892 Tau e.g.][]{2003A&A...402..701B}.  We estimate that we are sensitive to $\gtrsim$20 M$_{J}$ brown dwarf companions at L$^\prime$ band and $\gtrsim$50 M$_J$ companions at K$^\prime$ band, placing a rough upper limit on planetary masses in the system. We also convert the L$^\prime$ band contrast curve to planet mass times accretion rate \citep[e.g.][]{2015ApJ...803L...4E}. The L$^\prime$ band data are sensitive to a planet mass times accretion rate of $\sim$4 x $10^{-5}$ $\mathrm{M_J^{2}}$/yr, corresponding to a rapidly-accreting Jupiter analog or less-rapidly-accreting higher mass planet. 

\section{Discussion}\label{sec:discussion}
\subsection{System Geometry}\label{subsec:systemgeometry}
The geometry of the V892 Tau circumbinary disk measured here is consistent with the literature and we detect a new component of the disk structure with the discovery of the circumprimary. The diameter of the CB disk is in agreement with previous geometric constraints at $\sim$26 AU \citep[e.g.][]{2007ApJ...658.1164L,2008ApJ...681L..97M}. 
From the best-fit L$^\prime$ band model, we find that the position angle and orientation of the disk are similar to prior estimates made by \cite{2008ApJ...681L..97M} and \cite{2021ApJ...915..131L}. Previous inclination estimates ($i_{disk}=$ 54.6$^{\circ}$) indicate that the northwest side of the disk is closest to the observer \cite{2021ApJ...915..131L}. With this orientation, the inside of the disk rim on the southeast side of the star would be most visible to the observer.
This viewing angle effect, possibly combined with a puffed-up disk rim due to heating by the near-equal mass binary, may contribute to the asymmetry to the southeast of the star in the reconstructed images and geometric models.% The relatively low mass ratio of the binary suggests that the dominant asymmetry to the southeast of the star is an inclination effect rather than a physical disk rim asymmetry. 

 From the L’ reconstructed images, we find tentative evidence of warping in the circumbinary disk. In Section \ref{subsec:squeezemod} (Figure \ref{fig:modsqueezebinary}), we show that the Fourier observables from the geometric model cannot reproduce the complex asymmetry in the L$^\prime$ image reconstructed from the data (Figure \ref{fig:squeezebinary}). Near-equal mass binaries with highly eccentric orbits have been shown to cause warped circumbinary disks \citep[][]{{1994ApJ...421..651A,2020MNRAS.498.2936H,2015MNRAS.452.2396M}}. This scenario is consistent with  the V892 Tau eccentricity measurements in prior studies and this work (Table \ref{table:orbitresults}).

We further inform the architecture of the V892 Tau system with the first detection of a circumprimary disk with a diameter of $\sim$ 2 AU. Observations from \cite{2019PhDT.......134C} tentatively suggested a circumprimary disk, with differential phases in N band Mid-Infrared Interferometric Instrument (MIDI) data preferring either a circumprimary disk or an additional dusty companion. The K$^\prime$ band imaging presented here provides high enough angular resolution to firmly distinguish between these two physical interpretations. The rough alignment between the position angle of the companion and the skew angle of the circumprimary disk could suggest heating of the disk by the companion. Follow-up high angular resolution observations could identify whether the stellar companion causes the observed asymmetry by constraining the time evolution of the disk skew angle and the companion position angle. This work cannot place mass constraints on the circumprimary disk, but it nonetheless represents another potential reservoir of material for planet formation around V892 Tau.

\begin{deluxetable}{ccl}
\tablecaption{V892 Tau orbital parameters in comparison to \cite{2021ApJ...915..131L}. \label{table:orbitresults}}
\tablehead{
\colhead{Orbital parameter} & \colhead{Long et al. 2021} & \colhead{This work}} 
\startdata
Semi-major Axis (au)& 7.1 $\pm$ 0.1& 6.8 $\pm^{0.06}_{0.03}$  \\
Period (yrs)& 7.7 $\pm$ 0.2 & 7.2 $\pm^{0.04}_{0.06}$ \\
Eccentricity &0.27 $\pm$ 0.1& 0.25 $\pm$0.04\\
Inclination (degrees) & 59.3 $\pm$ 2.7 & 57.9 $\pm$2.8\\
Dynamical Mass (\(M_\odot\))& 6.0 $\pm$ 0.2 & 6.1 $\pm^{0.2}_{0.1}$\\
\enddata
\end{deluxetable}

\subsection{Orbit}
From the best-fit geometric models (discussed in Section \ref{sec:bestfit}), we measure the separation of the stellar companion to be 3.5 $\pm$ 0.1 AU at L$^\prime$ band. At K$^\prime$ band, we find the separation of the of the stellar companion is 5.6 AU $\pm$ 0.1 AU. 
The astrometric measurements are more precise at K$^\prime$ band than at L$^\prime$ band, since the angular resolution is higher due to the shorter wavelength. V892 Tau is also relatively bright at K$^\prime$ band (3.23 Jy at K$^\prime$ band and 1.75 Jy at L$^\prime$ band) and the K$^\prime$ band sky background is low. The high angular resolution, high signal-to-noise ratio, and AO correction with the PyWFS significantly reduces the size of error bars on the observables at both wavelengths, which are then propagated statistically through the best-fit astrometry and photometry.The eccentricity and inclination estimates are within $1\sigma$ of previously-published constraints, while the semi-major axis and period have a discrepancy of $2\sigma$ compared to those studies \citep{2021ApJ...915..131L}. We also find that our independent mass measurement of the system (6.1 +0.2/-0.1 $M_{\odot}$) is consistent with \citet{2021ApJ...915..131L}.

%include strehl values from JATIS paper (strehls for targets with v892 taus R band magnitude) this demonstrates that nrm and PYWFS is good for observing young stars 

%think about circumsecondary disk:
%calculate the expected K and L band flux ratio for a black body stellar atmosphere or do I need an excess at L band to explian the ratio of the fluxes
%flux at L may be a little high for just a stellar atmosphere
%cirumprimary disk is too small to resolve a L band; some extra dust may be around the comanpnion

\subsection{Benchmarking NRM with the PyWFS}
Since V892 Tau is bright at H band (the wavefront sensing bandpass), we see an improvement in contrast when compared to observations with the Shack-Hartmann wavefront sensor (SH WFS) at both science wavelengths, allowing us to make high-precision astrometric measurements. We compare the PyWFS contrast limits in Figure \ref{fig:cc} to contrast limits from \cite{2019JATIS...5a8001S} for a star with a similar R band (the SH WFS bandpass) magnitude to V892 Tau that was observed with Keck2/NIRC2 NRM behind the SH WFS. Like V892 Tau, this star is faint at R band, but bright at H band. 
%The Strehl ratio for the analogous star is 0.85 for L$'$ with the SH WFS. 
We find that the contrast is $\sim$0.5-1 magnitudes better with PyWFS than with the SH WFS. The boost in contrast demonstrates that the PyWFS is beneficial for observing red young stars. These observations are the first benchmark of NRM with Keck's PyWFS. Future observations of stars with fainter H band magnitudes will further quantify its performance in the lower-Strehl regime. 

\section{Conclusion}\label{sec:conclusion}
We presented multi-wavelength, multi-epoch Keck data of the V892 Tau circumbinary disk with NRM and the PyWFS. The data allow us to differentiate the secondary stellar emission from disk emission and deeply probe the structure of the disk at small angular separations. We fit geometric models to the L$^\prime$ and K$^\prime$ band data, and find that the morphologies of both images are best described by disk-plus-companion models. At L$^\prime$ band, the circumbinary disk properties are consistent with results from prior studies. At K$^\prime$ band, we make the first robust detection of a circumprimary disk. From the properties of the stellar binary, we update the orbit using our own data and archival data. This work places the tightest constraints on the orbital parameters of the V892 Tau stellar companion and the geometric structure of the circumbinary disk. Future observations of the V892 Tau system may include additional monitoring of the circumprimary disk to determine whether its skew is caused by heating from the stellar companion.

%From the statistics calculated in Table \ref{table:3}, we find that the disk-plus-companion model is %the only morphology that can adequately fit the squared visibility and closure phase data. At L band, the estimate in the error bars are conservative, which gives us reduced $\chi_{2}$ values that are $\leq$1. At K band, the reduced $\chi_{2}$ values have the closest value to 1 for the disk-plus-companion model with a circumprimary disk. When a circumbinary disk is forced, comparisons of the geometry to prior studies, and estimates within, are inconsistent. 

 These are the first published observations using NRM and the PyWFS in conjunction, providing a valuable contrast benchmark. We place mass constraints on undetected companions and compare the achieved contrast with the PyWFS to contrast limits achieved with the Shack-Hartmann WFS, finding a $\sim$0.5-1 magnitude boost in performance with the PyWFS. The exquisite AO correction (and thus achievable contrast) offered by the PyWFS enabled the precise astrometric measurements for V892 Tau, which improve on previous estimates by a factor of 10, and the high-angular-resolution detection of the circumprimary disk. These contrast benchmarks and the high-precision detections in the V892 Tau system demonstrate that future NRM surveys can take advantage of the PyWFS for observations of similarly young, red stars.

\acknowledgements
This material is based upon work supported by the National Science Foundation under Grant No. 2009698. J.A.E. acknowledges support from NSF award number 1745406. The data presented herein were obtained at the W. M. Keck Observatory, which is operated as a scientific partnership among the California Institute of Technology, the University of California and the National Aeronautics and Space Administration. The Observatory was made possible by the generous financial support of the W. M. Keck Foundation. The authors wish to recognize and acknowledge the very significant cultural role and reverence that the summit of Maunakea has always had within the indigenous Hawaiian community.  We are most fortunate to have the opportunity to conduct observations from this mountain.


\bibliography{sample63}{}

\begin{thebibliography}{}
\expandafter\ifx\csname natexlab\endcsname\relax\def\natexlab#1{#1}\fi
\providecommand{\url}[1]{\href{#1}{#1}}
\providecommand{\dodoi}[1]{doi:~\href{http://doi.org/#1}{\nolinkurl{#1}}}
\providecommand{\doeprint}[1]{\href{http://ascl.net/#1}{\nolinkurl{http://ascl.net/#1}}}
\providecommand{\doarXiv}[1]{\href{https://arxiv.org/abs/#1}{\nolinkurl{https://arxiv.org/abs/#1}}}

\bibitem[{{Akeson} {et~al.}(2013){Akeson}, {Chen}, {Ciardi}, {Crane}, {Good},
  {Harbut}, {Jackson}, {Kane}, {Laity}, {Leifer}, {Lynn}, {McElroy}, {Papin},
  {Plavchan}, {Ram{\'\i}rez}, {Rey}, {von Braun}, {Wittman}, {Abajian}, {Ali},
  {Beichman}, {Beekley}, {Berriman}, {Berukoff}, {Bryden}, {Chan}, {Groom},
  {Lau}, {Payne}, {Regelson}, {Saucedo}, {Schmitz}, {Stauffer}, {Wyatt}, \&
  {Zhang}}]{2013PASP..125..989A}
{Akeson}, R.~L., {Chen}, X., {Ciardi}, D., {et~al.} 2013, \pasp, 125, 989,
  \dodoi{10.1086/672273}

\bibitem[{{Artymowicz} \& {Lubow}(1994)}]{1994ApJ...421..651A}
{Artymowicz}, P., \& {Lubow}, S.~H. 1994, \apj, 421, 651,
  \dodoi{10.1086/173679}

\bibitem[{{Baldwin} {et~al.}(1986){Baldwin}, {Haniff}, {Mackay}, \&
  {Warner}}]{1986Natur.320..595B}
{Baldwin}, J.~E., {Haniff}, C.~A., {Mackay}, C.~D., \& {Warner}, P.~J. 1986,
  \nat, 320, 595, \dodoi{10.1038/320595a0}

\bibitem[{{Baraffe} {et~al.}(2003){Baraffe}, {Chabrier}, {Barman}, {Allard}, \&
  {Hauschildt}}]{2003A&A...402..701B}
{Baraffe}, I., {Chabrier}, G., {Barman}, T.~S., {Allard}, F., \& {Hauschildt},
  P.~H. 2003, \aap, 402, 701, \dodoi{10.1051/0004-6361:20030252}

\bibitem[{{Baron} {et~al.}(2010){Baron}, {Monnier}, \&
  {Kloppenborg}}]{2010SPIE.7734E..2IB}
{Baron}, F., {Monnier}, J.~D., \& {Kloppenborg}, B. 2010, in Society of
  Photo-Optical Instrumentation Engineers (SPIE) Conference Series, Vol. 7734,
  Optical and Infrared Interferometry II, ed. W.~C. {Danchi}, F.~{Delplancke},
  \& J.~K. {Rajagopal}, 77342I, \dodoi{10.1117/12.857364}

\bibitem[{{Benest}(1993)}]{1993CeMDA..56...45B}
{Benest}, D. 1993, Celestial Mechanics and Dynamical Astronomy, 56, 45,
  \dodoi{10.1007/BF00699718}

\bibitem[{{Blunt} {et~al.}(2020){Blunt}, {Wang}, {Angelo}, {Ngo}, {Cody}, {De
  Rosa}, {Graham}, {Hirsch}, {Nagpal}, {Nielsen}, {Pearce}, {Rice}, \&
  {Tejada}}]{2020AJ....159...89B}
{Blunt}, S., {Wang}, J.~J., {Angelo}, I., {et~al.} 2020, \aj, 159, 89,
  \dodoi{10.3847/1538-3881/ab6663}

\bibitem[{{Boehler} {et~al.}(2017){Boehler}, {Weaver}, {Isella}, {Ricci},
  {Grady}, {Carpenter}, \& {Perez}}]{2017ApJ...840...60B}
{Boehler}, Y., {Weaver}, E., {Isella}, A., {et~al.} 2017, \apj, 840, 60,
  \dodoi{10.3847/1538-4357/aa696c}

\bibitem[{{Cahuasqu{\'\i}}(2019)}]{2019PhDT.......134C}
{Cahuasqu{\'\i}}, J.~A. 2019, PhD thesis, Andreas Eckart University of Cologne,
  Germany

\bibitem[{{Coleman} {et~al.}(2022){Coleman}, {Nelson}, \&
  {Triaud}}]{2022MNRAS.513.2563C}
{Coleman}, G. A.~L., {Nelson}, R.~P., \& {Triaud}, A. H.~M.~J. 2022, \mnras,
  513, 2563, \dodoi{10.1093/mnras/stac1029}

\bibitem[{{Doyle} {et~al.}(2011){Doyle}, {Carter}, {Fabrycky}, {Slawson},
  {Howell}, {Winn}, {Orosz}, {P{\v{r}}sa}, {Welsh}, {Quinn}, {Latham},
  {Torres}, {Buchhave}, {Marcy}, {Fortney}, {Shporer}, {Ford}, {Lissauer},
  {Ragozzine}, {Rucker}, {Batalha}, {Jenkins}, {Borucki}, {Koch}, {Middour},
  {Hall}, {McCauliff}, {Fanelli}, {Quintana}, {Holman}, {Caldwell}, {Still},
  {Stefanik}, {Brown}, {Esquerdo}, {Tang}, {Furesz}, {Geary}, {Berlind},
  {Calkins}, {Short}, {Steffen}, {Sasselov}, {Dunham}, {Cochran}, {Boss},
  {Haas}, {Buzasi}, \& {Fischer}}]{2011Sci...333.1602D}
{Doyle}, L.~R., {Carter}, J.~A., {Fabrycky}, D.~C., {et~al.} 2011, Science,
  333, 1602, \dodoi{10.1126/science.1210923}

\bibitem[{{Eisner}(2015)}]{2015ApJ...803L...4E}
{Eisner}, J.~A. 2015, \apjl, 803, L4, \dodoi{10.1088/2041-8205/803/1/L4}

\bibitem[{{El-Badry} {et~al.}(2021){El-Badry}, {Rix}, \&
  {Heintz}}]{2021MNRAS.506.2269E}
{El-Badry}, K., {Rix}, H.-W., \& {Heintz}, T.~M. 2021, \mnras, 506, 2269,
  \dodoi{10.1093/mnras/stab323}

\bibitem[{Foreman-Mackey {et~al.}(2013)Foreman-Mackey, Hogg, Lang, \&
  Goodman}]{Foreman_Mackey_2013}
Foreman-Mackey, D., Hogg, D.~W., Lang, D., \& Goodman, J. 2013, Publications of
  the Astronomical Society of the Pacific, 125, 306, \dodoi{10.1086/670067}

\bibitem[{{Gaia Collaboration}(2020)}]{2020yCat.1350....0G}
{Gaia Collaboration}. 2020, VizieR Online Data Catalog, I/350

\bibitem[{{Guilloteau} {et~al.}(2008){Guilloteau}, {Dutrey}, {Pety}, \&
  {Gueth}}]{2008A&A...478L..31G}
{Guilloteau}, S., {Dutrey}, A., {Pety}, J., \& {Gueth}, F. 2008, \aap, 478,
  L31, \dodoi{10.1051/0004-6361:20079053}

\bibitem[{{Guyon} {et~al.}(2006){Guyon}, {Pluzhnik}, {Kuchner}, {Collins}, \&
  {Ridgway}}]{2006ApJS..167...81G}
{Guyon}, O., {Pluzhnik}, E.~A., {Kuchner}, M.~J., {Collins}, B., \& {Ridgway},
  S.~T. 2006, \apjs, 167, 81, \dodoi{10.1086/507630}

\bibitem[{{Guyon} {et~al.}(2013){Guyon}, {Eisner}, {Angel}, {Woolf}, {Bendek},
  {Milster}, {Ammons}, {Shao}, {Shaklan}, {Levine}, {Nemati}, {Martinache},
  {Pitman}, {Woodruff}, \& {Belikov}}]{2013ApJ...767...11G}
{Guyon}, O., {Eisner}, J.~A., {Angel}, R., {et~al.} 2013, \apj, 767, 11,
  \dodoi{10.1088/0004-637X/767/1/11}

\bibitem[{{Herczeg} \& {Hillenbrand}(2014)}]{2014ApJ...786...97H}
{Herczeg}, G.~J., \& {Hillenbrand}, L.~A. 2014, \apj, 786, 97,
  \dodoi{10.1088/0004-637X/786/2/97}

\bibitem[{{Hern{\'a}ndez} {et~al.}(2004){Hern{\'a}ndez}, {Calvet},
  {Brice{\~n}o}, {Hartmann}, \& {Berlind}}]{2004AJ....127.1682H}
{Hern{\'a}ndez}, J., {Calvet}, N., {Brice{\~n}o}, C., {Hartmann}, L., \&
  {Berlind}, P. 2004, \aj, 127, 1682, \dodoi{10.1086/381908}

\bibitem[{{Hillenbrand} {et~al.}(1992){Hillenbrand}, {Strom}, {Vrba}, \&
  {Keene}}]{1992ApJ...397..613H}
{Hillenbrand}, L.~A., {Strom}, S.~E., {Vrba}, F.~J., \& {Keene}, J. 1992, \apj,
  397, 613, \dodoi{10.1086/171819}

\bibitem[{{Hirsh} {et~al.}(2020){Hirsh}, {Price}, {Gonzalez},
  {Ubeira-Gabellini}, \& {Ragusa}}]{2020MNRAS.498.2936H}
{Hirsh}, K., {Price}, D.~J., {Gonzalez}, J.-F., {Ubeira-Gabellini}, M.~G., \&
  {Ragusa}, E. 2020, \mnras, 498, 2936, \dodoi{10.1093/mnras/staa2536}

\bibitem[{{Ireland}(2013)}]{2013MNRAS.433.1718I}
{Ireland}, M.~J. 2013, \mnras, 433, 1718, \dodoi{10.1093/mnras/stt859}

\bibitem[{{Ireland} \& {Kraus}(2008)}]{2008ApJ...678L..59I}
{Ireland}, M.~J., \& {Kraus}, A.~L. 2008, \apjl, 678, L59,
  \dodoi{10.1086/588216}

\bibitem[{{Jennison}(1958)}]{1958MNRAS.118..276J}
{Jennison}, R.~C. 1958, \mnras, 118, 276, \dodoi{10.1093/mnras/118.3.276}

\bibitem[{{Kley} \& {Haghighipour}(2014)}]{2014A&A...564A..72K}
{Kley}, W., \& {Haghighipour}, N. 2014, \aap, 564, A72,
  \dodoi{10.1051/0004-6361/201323235}

\bibitem[{{Kurtovic} {et~al.}(2018){Kurtovic}, {P{\'e}rez}, {Benisty}, {Zhu},
  {Zhang}, {Huang}, {Andrews}, {Dullemond}, {Isella}, {Bai}, {Carpenter},
  {Guzm{\'a}n}, {Ricci}, \& {Wilner}}]{2018ApJ...869L..44K}
{Kurtovic}, N.~T., {P{\'e}rez}, L.~M., {Benisty}, M., {et~al.} 2018, \apjl,
  869, L44, \dodoi{10.3847/2041-8213/aaf746}

\bibitem[{{Lines} {et~al.}(2015){Lines}, {Leinhardt}, {Baruteau},
  {Paardekooper}, \& {Carter}}]{2015A&A...582A...5L}
{Lines}, S., {Leinhardt}, Z.~M., {Baruteau}, C., {Paardekooper}, S.~J., \&
  {Carter}, P.~J. 2015, \aap, 582, A5, \dodoi{10.1051/0004-6361/201526295}

\bibitem[{{Liu} {et~al.}(2007){Liu}, {Hinz}, {Meyer}, {Mamajek}, {Hoffmann},
  {Brusa}, {Miller}, \& {Kenworthy}}]{2007ApJ...658.1164L}
{Liu}, W.~M., {Hinz}, P.~M., {Meyer}, M.~R., {et~al.} 2007, \apj, 658, 1164,
  \dodoi{10.1086/511779}

\bibitem[{{Long} {et~al.}(2021){Long}, {Andrews}, {Vega}, {Wilner}, {Chandler},
  {Ragusa}, {Teague}, {P{\'e}rez}, {Calvet}, {Carpenter}, {Henning}, {Kwon},
  {Linz}, \& {Ricci}}]{2021ApJ...915..131L}
{Long}, F., {Andrews}, S.~M., {Vega}, J., {et~al.} 2021, \apj, 915, 131,
  \dodoi{10.3847/1538-4357/abff53}

\bibitem[{{Masset} {et~al.}(2006){Masset}, {Morbidelli}, {Crida}, \&
  {Ferreira}}]{2006ApJ...642..478M}
{Masset}, F.~S., {Morbidelli}, A., {Crida}, A., \& {Ferreira}, J. 2006, \apj,
  642, 478, \dodoi{10.1086/500967}

\bibitem[{{Mawet} {et~al.}(2012){Mawet}, {Pueyo}, {Lawson}, {Mugnier}, {Traub},
  {Boccaletti}, {Trauger}, {Gladysz}, {Serabyn}, {Milli}, {Belikov}, {Kasper},
  {Baudoz}, {Macintosh}, {Marois}, {Oppenheimer}, {Barrett}, {Beuzit},
  {Devaney}, {Girard}, {Guyon}, {Krist}, {Mennesson}, {Mouillet}, {Murakami},
  {Poyneer}, {Savransky}, {V{\'e}rinaud}, \& {Wallace}}]{2012SPIE.8442E..04M}
{Mawet}, D., {Pueyo}, L., {Lawson}, P., {et~al.} 2012, in \procspie, Vol. 8442,
  Space Telescopes and Instrumentation 2012: Optical, Infrared, and Millimeter
  Wave, 844204, \dodoi{10.1117/12.927245}

\bibitem[{{Miranda} \& {Lai}(2015)}]{2015MNRAS.452.2396M}
{Miranda}, R., \& {Lai}, D. 2015, \mnras, 452, 2396,
  \dodoi{10.1093/mnras/stv1450}

\bibitem[{{Monnier} {et~al.}(2008){Monnier}, {Tannirkulam}, {Tuthill},
  {Ireland}, {Cohen}, {Danchi}, \& {Baron}}]{2008ApJ...681L..97M}
{Monnier}, J.~D., {Tannirkulam}, A., {Tuthill}, P.~G., {et~al.} 2008, \apjl,
  681, L97, \dodoi{10.1086/590532}

\bibitem[{{Monnier} {et~al.}(2009){Monnier}, {Tuthill}, {Ireland}, {Cohen},
  {Tannirkulam}, \& {Perrin}}]{2009ApJ...700..491M}
{Monnier}, J.~D., {Tuthill}, P.~G., {Ireland}, M., {et~al.} 2009, \apj, 700,
  491, \dodoi{10.1088/0004-637X/700/1/491}

\bibitem[{{M{\"u}ller} \& {Kley}(2012)}]{2012A&A...539A..18M}
{M{\"u}ller}, T.~W.~A., \& {Kley}, W. 2012, \aap, 539, A18,
  \dodoi{10.1051/0004-6361/201118202}

\bibitem[{{Penzlin} {et~al.}(2021){Penzlin}, {Kley}, \&
  {Nelson}}]{2021A&A...645A..68P}
{Penzlin}, A. B.~T., {Kley}, W., \& {Nelson}, R.~P. 2021, \aap, 645, A68,
  \dodoi{10.1051/0004-6361/202039319}

\bibitem[{{Pierens} \& {Nelson}(2008)}]{2008A&A...483..633P}
{Pierens}, A., \& {Nelson}, R.~P. 2008, \aap, 483, 633,
  \dodoi{10.1051/0004-6361:200809453}

\bibitem[{{Rabl} \& {Dvorak}(1988)}]{1988A&A...191..385R}
{Rabl}, G., \& {Dvorak}, R. 1988, \aap, 191, 385

\bibitem[{{Ruane} {et~al.}(2019){Ruane}, {Mawet}, {Riggs}, \&
  {Serabyn}}]{2019SPIE11117E..1FR}
{Ruane}, G., {Mawet}, D., {Riggs}, A.~J.~E., \& {Serabyn}, E. 2019, in Society
  of Photo-Optical Instrumentation Engineers (SPIE) Conference Series, Vol.
  11117, Society of Photo-Optical Instrumentation Engineers (SPIE) Conference
  Series, 111171F, \dodoi{10.1117/12.2528625}

\bibitem[{{Sallum} \& {Eisner}(2017)}]{2017ApJS..233....9S}
{Sallum}, S., \& {Eisner}, J. 2017, \apjs, 233, 9,
  \dodoi{10.3847/1538-4365/aa90bb}

\bibitem[{{Sallum} {et~al.}(2021){Sallum}, {Eisner}, {Stone}, {Dietrich},
  {Hinz}, \& {Spalding}}]{2021AJ....161...28S}
{Sallum}, S., {Eisner}, J.~A., {Stone}, J.~M., {et~al.} 2021, \aj, 161, 28,
  \dodoi{10.3847/1538-3881/abc957}

\bibitem[{{Sallum} {et~al.}(2022){Sallum}, {Ray}, \&
  {Hinkley}}]{2022SPIE12183E..2MS}
{Sallum}, S., {Ray}, S., \& {Hinkley}, S. 2022, in Society of Photo-Optical
  Instrumentation Engineers (SPIE) Conference Series, Vol. 12183, Optical and
  Infrared Interferometry and Imaging VIII, ed. A.~{M{\'e}rand}, S.~{Sallum},
  \& J.~{Sanchez-Bermudez}, 121832M, \dodoi{10.1117/12.2630401}

\bibitem[{{Sallum} \& {Skemer}(2019)}]{2019JATIS...5a8001S}
{Sallum}, S., \& {Skemer}, A. 2019, Journal of Astronomical Telescopes,
  Instruments, and Systems, 5, 018001, \dodoi{10.1117/1.JATIS.5.1.018001}

\bibitem[{{Sallum} {et~al.}(2019){Sallum}, {Skemer}, {Eisner}, {van der Marel},
  {Sheehan}, {Close}, {Ireland}, {Males}, {Morzinski}, {Bailey}, {Briguglio},
  \& {Puglisi}}]{2019ApJ...883..100S}
{Sallum}, S., {Skemer}, A.~J., {Eisner}, J.~A., {et~al.} 2019, \apj, 883, 100,
  \dodoi{10.3847/1538-4357/ab3dae}

\bibitem[{{Sigurdsson} {et~al.}(2003){Sigurdsson}, {Richer}, {Hansen},
  {Stairs}, \& {Thorsett}}]{2003Sci...301..193S}
{Sigurdsson}, S., {Richer}, H.~B., {Hansen}, B.~M., {Stairs}, I.~H., \&
  {Thorsett}, S.~E. 2003, Science, 301, 193, \dodoi{10.1126/science.1086326}

\bibitem[{{Smith} {et~al.}(2005){Smith}, {Balega}, {Duschl}, {Hofmann},
  {Lachaume}, {Preibisch}, {Schertl}, \& {Weigelt}}]{2005A&A...431..307S}
{Smith}, K.~W., {Balega}, Y.~Y., {Duschl}, W.~J., {et~al.} 2005, \aap, 431,
  307, \dodoi{10.1051/0004-6361:20041135}

\bibitem[{{Socia} {et~al.}(2020){Socia}, {Welsh}, {Orosz}, {Cochran}, {Endl},
  {Quarles}, {Short}, {Torres}, {Windmiller}, \&
  {Yenawine}}]{2020AJ....159...94S}
{Socia}, Q.~J., {Welsh}, W.~F., {Orosz}, J.~A., {et~al.} 2020, \aj, 159, 94,
  \dodoi{10.3847/1538-3881/ab665b}

\bibitem[{{Strom} \& {Strom}(1994)}]{1994ApJ...424..237S}
{Strom}, K.~M., \& {Strom}, S.~E. 1994, \apj, 424, 237, \dodoi{10.1086/173886}

\bibitem[{{Szebehely}(1980)}]{1980CeMec..22....7S}
{Szebehely}, V. 1980, Celestial Mechanics, 22, 7, \dodoi{10.1007/BF01228750}

\bibitem[{{Tuthill} {et~al.}(2000){Tuthill}, {Monnier}, {Danchi}, {Wishnow}, \&
  {Haniff}}]{2000PASP..112..555T}
{Tuthill}, P.~G., {Monnier}, J.~D., {Danchi}, W.~C., {Wishnow}, E.~H., \&
  {Haniff}, C.~A. 2000, \pasp, 112, 555, \dodoi{10.1086/316550}

\bibitem[{{Zhu}(2015)}]{2015ApJ...799...16Z}
{Zhu}, Z. 2015, \apj, 799, 16, \dodoi{10.1088/0004-637X/799/1/16}

\end{thebibliography}

\bibliographystyle{aasjournal}


\end{document}